\documentclass[num-refs]{wiley-article}

\usepackage{siunitx}
\usepackage{upgreek} 
\usepackage{color} 
\definecolor{blue}{rgb}{0,0,0.8}
\definecolor{fuchsia}{rgb}{0.5,0,0.8}

\definecolor{red}{rgb}{0.8,0,0}

\definecolor{green}{rgb}{0,0.4,0}

\usepackage{wrapfig}
\usepackage{floatrow}
\newfloatcommand{capbtabbox}{table}[][\FBwidth]
\usepackage{hyperref}
\usepackage{ulem}
\usepackage{amsmath}
\usepackage{multirow}
\usepackage{multicol}

\papertype{Research Article}

\title{Double Diffusion Encoding Prevents Degeneracy in Parameter Estimation of Biophysical Models in Diffusion {MRI}}

\abbrevs{dMRI, diffusion MRI; SM, Standard Model; NODDI(DA), neurite orientation dispersion and density imaging (with diffusivity assessment); SNR, signal-to-noise ratio; 3D/5D, three/five dimensional; SDE, single diffusion encoding; DDE, double diffusion encoding; LTE, linear tensor encoding; PTE, planar tensor encoding; STE, spherical tensor encoding; (R)MSE, (root) mean squared error.}

\author[1,2]{Santiago Coelho}
\author[1,2]{Jose M. Pozo}
\author[3,4]{Sune N. Jespersen}
\author[5,6]{Derek K. Jones}
\author[1,2]{Alejandro F. Frangi}

\affil[1]{Centre for Computational Imaging \& Simulation Technologies in Biomedicine (CISTIB) and Leeds Institute for Cardiac and Metabolic Medicine (LICAMM), School of Computing \& School of Medicine, University of Leeds, Leeds,  UK}
\affil[2]{CISTIB, Electronic and Electrical Engineering Department, The University of Sheffield, Sheffield, UK}
\affil[3]{Center of Functionally Integrative Neuroscience (CFIN) and MINDLab, Department of Clinical Medicine, Aarhus University, Aarhus, Denmark}
\affil[4]{Department of Physics and Astronomy, Aarhus University, Aarhus, Denmark}
\affil[5]{Cardiff University Brain Research Imaging Centre (CUBRIC), Cardiff University, UK}
\affil[6]{School of Psychology, Australian Catholic University, Melbourne, Australia.}

\corraddress{Santiago Coelho, CISTIB, School of Computing, University of Leeds, EC Stoner Building, Rm 6.08, Woodhouse Lane, Leeds LS2 9JT, UK.}
\corremail{s.coelho@leeds.ac.uk}

\usepackage{fancyhdr}
\runningauthor{Santiago Coelho et al.}

\begin{document}

\maketitle


\begin{abstract}
\subsection*{Purpose}
Biophysical tissue models are increasingly used in the interpretation of diffusion MRI (dMRI) data, with the potential to provide specific biomarkers of brain microstructural changes. However, the general Standard Model has recently shown that model parameter estimation from dMRI data is ill-posed unless very strong magnetic gradients are used. We analyse this issue for the Neurite Orientation Dispersion and Density Imaging with Diffusivity Assessment (NODDIDA) model and demonstrate that its extension from Single Diffusion Encoding (SDE) to Double Diffusion Encoding (DDE) solves the ill-posedness and increases the accuracy of the parameter estimation.
\subsection*{Methods}
We analyse theoretically the cumulant expansion up to fourth order in b of SDE and DDE signals.
Additionally, we perform \textit{in silico} experiments to compare SDE and DDE capabilities under similar noise conditions.
\subsection*{Results}
We prove analytically that DDE provides invariant information non-accessible from SDE, which makes the NODDIDA parameter estimation injective. The \textit{in silico} experiments show that DDE reduces the bias and mean square error of the estimation along the whole feasible region of 5D model parameter space.
\subsection*{Conclusions}
DDE adds additional information for estimating the model parameters, unexplored by SDE, which is enough to solve the degeneracy in the NODDIDA model parameter estimation.



\end{abstract}



\section{Introduction}

Diffusion MRI (dMRI) has been established as an invaluable tool for characterising brain microstructure \textit{in vivo} and non-invasively. 
Diffusion weighted images (DWIs) are sensitive to the random displacement of water molecules within a voxel \cite{CALLAGHAN2010}, 
probing tissue on scales considerably lower than image resolution \cite{KISELEV2016}.
Diffusion MRI provides the aggregate signal from the distribution of components within a voxel. By measuring across multiple diffusion orientations and weightings, information about the underlying tissue architecture can be unravelled. The ability to detect small alterations in brain tissue is a key factor when developing biomarkers for early stages of neurodegenerative diseases \cite{ASSAF2008b}. Various approaches to derive information from Diffusion Weighted Images (DWI) have been proposed in the literature \cite{BASSER1994,ASSAF1999,TUCH2004,TOURNIER2004,JENSEN2005}. Most direct approaches, such as Diffusion Tensor Imaging (DTI) \cite{BASSER1994}, are just aimed at describing the main MRI signal characteristics (signal representations, \cite{NOVIKOV2018b}). However, the quest for specific information on tissue microstructural integrity inspired the development of biophysical tissue models \cite{VANGELDEREN1994,STANISZ1997,ASSAF2004a,JESPERSEN2007}.
By assuming certain characteristics on tissue properties (such as their geometry) these models allow the extraction of more specific microstructural information than signal representations. Nevertheless, the validity of these results relies on how accurate the model is for the tissue under study. The widely used Neurite Orientation Dispersion and Density Imaging (NODDI) \cite{ZHANG2012} model fixes the diffusivity values of the compartments present in the voxel to specific values. This assumption has been challenged in \cite{LAMPINEN2017a} and it has been argued to introduce bias in the estimation of the remaining model parameters \cite{HUTCHINSON2017}. To overcome this limitation, Jelescu \textit{et al.} \cite{JELESCU2015a} extended the model by adding the diffusivities to the estimation routine (they dubbed it NODDIDA, NODDI with Diffusivity Assessment). While this approach 
eliminated some flawed assumptions made by NODDI, this led to multiple possible solutions that describe the signal equally well. This reflects that the estimation problem is ill-posed or, at least, ill-conditioned, and is usually stated as the existence of degenerated model parameter sets. Recent work by Novikov \textit{et al.} showed that this degeneracy is intrinsic to the so-called standard model \cite{NOVIKOV2018}, and that one must employ high b-values to overcome it. Furthermore, Reisert \textit{et al.} \cite{REISERT2017} proposed a supervised machine learning approach to circumvent the degeneracy in the parameter estimation.
%
%
%
%

Most of the dMRI techniques have been developed for an acquisition performed within a Single Diffusion Encoding (SDE) framework. 
%
%
%
Since Stejskal and Tanner developed the Pulsed Gradient Spin Echo (PGSE) sequence \cite{STEJSKAL&TANNER1965}, there have been many works aimed at maximising the information that can be obtained from a dMRI experiment
by exploring different acquisition protocols \cite{JONES2004,ALEXANDER2008}. One of the many modifications proposed to the magnetic gradient waveforms involves the addition of multiple gradient pairs. Particularly, a scheme that has lately gained popularity is termed Double Diffusion Encoding (DDE) \cite{SHEMESH2015}, first proposed by Cory \textit{et al.} \cite{CORY1990}. Analogously to SDE, the term DDE refers to any sequence consisting of two consecutive diffusion encodings.
It has been shown that DDE has the potential to provide new information that is not immediately accessible with SDE \cite{SHEMESH2010a}. Many groups focused on developing methods for extracting microstructural information based on this scheme \cite{OZARSLAN2009a,JESPERSEN2013,BENJAMINI2014,IANUS2016}. Jespersen \textit{et al.} \cite{JESPERSEN2011} showed that in the low-diffusion-weighting limit, the information extracted from single and multiple diffusion encodings is the same. Recently, Lampinen \textit{et al.} \cite{LAMPINEN2017a} have analysed the advantages of a 
multidimensional encoding over SDE NODDI. They proved that extending the acquisition 
increases the accuracy in quantifying microscopic anisotropy. However, it has not been fully explored, from the point of view of fitting a biophysical model to noisy measurements, if single or multiple encodings can provide us with more precise model parameter estimates (\textit{cf.} \cite{BENJAMINI2014,IANUS2016}). Recently,
the advantages of combining linear with planar or spherical tensor encoding to lift the degeneracy and increase the parameter estimation precision have been investigated \cite{COELHO2017,FIEREMANS2018,DHITAL2018} through \textit{in silico} experiments. 
Their results show that the estimation precision is increased by the addition of 
these orthogonal measurements. However, a theoretical background of why this happens is still missing.
%
%



This paper extends NODDIDA to a DDE scheme and assesses the accuracy of estimators based on SDE and DDE measurements. This extension adds more degrees of freedom to the data acquisition (\textit{i.e.} two diffusion encoding periods must be chosen). 
We hypothesised that DDE acquisition protocols containing both parallel and perpendicular direction pairs might outperform SDE protocols in informing biophysical models. 
%
%
%
%
%
We investigated analytically the different information provided by DDE and SDE 
in terms of their 4th order cumulant expansions. 
We examine the ill-posedness of the parameter estimation from SDE and present a theoretical explanation of why DDE resolves the degeneracy without reaching extremely high diffusion weightings (\textit{e.g.} $b > 4 ms/\mu m^2$). Additionally, we generated \textit{in silico} dMRI measurements for acquisitions with different DDE configurations from a wide range of model parameter values covering the biologically feasible region of the 5D parameter space. Under similar experimental conditions, the higher accuracy is obtained for DDE combining parallel and perpendicular direction pairs, outperforming SDE in most scenarios.


\newpage

\section{Theory}\label{S:THEORY}
\subsection{Biophysical model assumptions}

A general assumption among multi-compartment models representing tissue microstructure is that water exchange between compartments is negligible for typical experimental time scales. The total signal is the weighted contribution from each compartment. The two-compartment model dubbed {\it Standard Model} (SM) is the most general version of the typical models used for diffusion in neuronal tissue (see \cite{NOVIKOV2018}). 
The intra-neurite compartment represents axons and glial processes with restricted diffusion (\textit{e.g.} \cite{JESPERSEN2007}). This is modelled as narrow `sticks', where diffusion is assumed to occur only 
along the fibre's main direction. The extra-neurite compartment is considered to have a hindered diffusion, modelled as anisotropic (\textit{e.g.} \cite{JESPERSEN2010}).
A \textit{fibre segment} is defined as the local bundle of aligned axons and astrocyte processes with the extra-neurite space surrounding them. Voxels are composed of a large number of fibre segments. The SM consists of the fibre segment signal model (\textit{i.e.} kernel) with the diffusivities and water fraction as free parameters, together with a general fibre orientation distribution function (ODF), which could be represented by its spherical harmonics decomposition.

Some other works consider a third compartment that represents the contribution from stationary water \cite{STANISZ1997,ALEXANDER2010}. However, recent works \cite{DHITAL2017} have concluded that the signal arising from this compartment can be neglected in most structures for the diffusion times used in the clinic and should only be considered in the cerebellum \cite{TAX2018}.

\begin{figure}[htbp]
\caption{Diagram of the two compartments present in the NODDIDA tissue model with their corresponding diffusivities.}\label{fig:TwoCompartments}
\includegraphics[scale=.23]{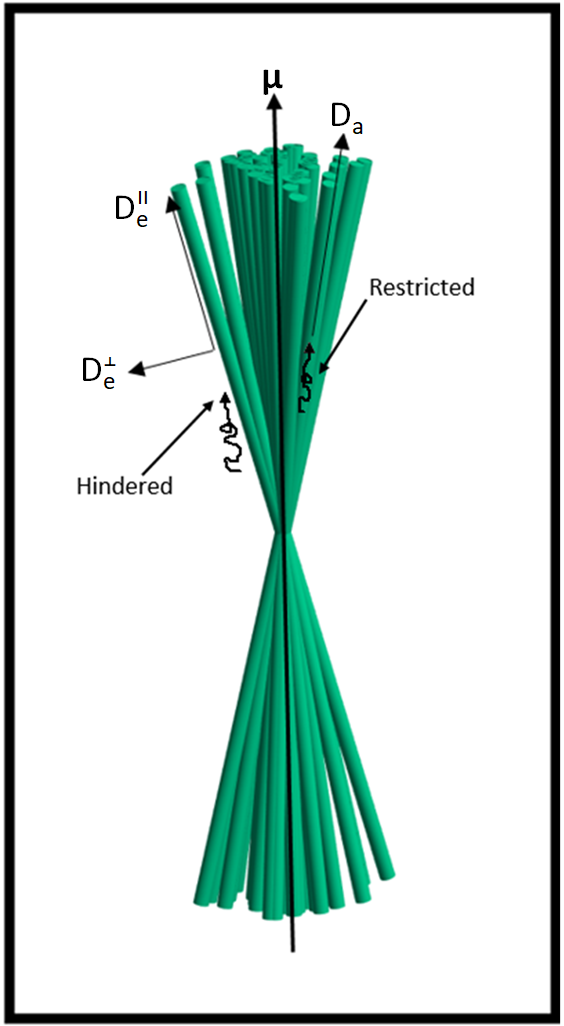}
\end{figure}

Considering a general fibre ODF involves a large set of parameters, which can hinder their unambiguous estimation from the dMRI signal. The NODDIDA model \cite{JELESCU2015a}, is essentially the SM with the constraint that the fibre ODF must be a Watson spherical distribution $\mathcal{P}(\hat{\mathbf{u}}) = f(\hat{\mathbf{u}} \mid \hat{\pmb{\mu}},\kappa)$, with concentration parameter $\kappa$ and main direction $\hat{\pmb{\mu}}$  (see Fig. \ref{fig:TwoCompartments}). This cylindrically symmetric ODF is usually considered a good model, especially for white matter regions without crossing fibres. Although being a simplified version of SM, NODDIDA still presents some degeneracy problems. Thus,  in this work, we focus our analysis on the NODIDDA model.

\subsection{NODDIDA model with SDE}
For a general SM, the signal from a SDE experiment, where the diffusion weighting $b$ 
(\textit{i.e.} b-value)
is applied in the diffusion encoding direction $\hat{\mathbf{n}}= [n_x, n_y, n_z]^t$, is given by the convolution over the unit sphere \cite{NOVIKOV2018}
 \begin{equation}\label{eq:IntegratedKernel}
 	S_{\text{SDE}}(b,\hat{\mathbf{n}})= 
         S_0 \int_{\mathbb{S}^2}\mathcal{P}(\hat{\mathbf{u}})
        \mathcal{K}(b,\hat{\mathbf{n}}\cdot\hat{\mathbf{u}}) \,
     dS_{\hat{\mathbf{u}}},
 \end{equation}
where 
\begin{equation}\label{eq:Kernel_SDE}
	\mathcal{K}(b,\hat{\mathbf{n}}\cdot\hat{\mathbf{u}}) = 
    	f \exp\bigl[-b D_{\text{a}}
        	(\hat{\mathbf{n}}\cdot\hat{\mathbf{u}})^2\bigr] +
        (1-f) \exp\bigl[-b D_\text{e}^\perp - 
        	b \Delta_\text{e} 
            (\hat{\mathbf{n}}\cdot\hat{\mathbf{u}})^2\bigr],
\end{equation}
is the response signal (kernel) from a fibre segment oriented along direction $\hat{\mathbf{u}}$. Here, $f$ is the $T_2$-weighted intra-neurite volume fraction, $D_{\text{a}}$ the intra-neurite diffusivity, and $\Delta_\text{e}=D_\text{e}^\parallel-D_\text{e}^\perp$, with $D_\text{e}^\parallel$, $D_\text{e}^\perp$ the extra-neurite diffusivities parallel and perpendicular to the fibre-segment axis. These scalar kernel parameters ($f$, $D_{\text{a}}$, $D_\text{e}^\parallel$, and $D_\text{e}^\perp$) provide important tissue microstructural information, and have shown potential clinical relevance as they are sensitive to specific disease processes such as demyelination, axonal loss or inflammation \cite{FIEREMANS2012,JELESCU2016,BUDDE2010}.

%
%

It has been recently shown that the parameter estimation is challenging under normal experimental conditions. There are two issues here. The first one is that fitting these models to noisy measurements is generally a non-convex optimisation problem, potentially having several local minima of the objective function,
requiring appropriate optimisation algorithms. However, the existence of multiple local minima opens the door to a second, more serious, issue: the objective function can present multiple minima with equal or very similar values. 
In the presence of noise these minima are perturbed, 
making unstable which one becomes the global minimum.
Jelescu \textit{et al.} \cite{JELESCU2015b} evidenced this ill-posedness issue for clinically feasible dMRI acquisitions in two particular cases. 
They showed that the estimated parameters from a collection of independently simulated dMRI measurements
follow a bi-modal distribution, despite being simulated from a single ground truth, and the presence of practically indistinguishable spurious minima in the objective function. 

\subsection{Parameter estimation from SDE: an ill-posed problem}\label{S:SDE_ill_posed_problem}
A recent work by Novikov \textit{et al.} \cite{NOVIKOV2018} analysed in detail this inverse problem 
for the unconstrained SM by reparametrising it into its rotational invariants. 
They concluded that without any constraints on the ODF shape, it was not possible to estimate the kernel parameters with an acquisition sensitive up to order $\mathcal{O}(b^2)$. However, in this work we are interested in studying NODDIDA, where the ODF is given by a Watson distribution. 
%
%
%
%
%

For intermediate diffusion weightings (\textit{i.e.} $b<2.5ms/\mu m^2$) the dMRI signal is accurately represented by its $4^{th}$-order cumulant expansion (sensitive up to $\mathcal{O}(b^2)$ contributions). For SDE this expansion can be written as  \cite{JENSEN2005}
\begin{equation}\label{Eq:Cumulant4thOrder_SDE}
	\log (S(b,\hat{\mathbf{n}})/S_0)  \approx 
    - b n_i n_j D_{ij} + 
    \frac{1}{6} b^2 \bar{D}^2 
    	n_i n_j n_k n_\ell W_{ijk\ell} = 
    - b \textbf{D}(\hat{\mathbf{n}}) + 
    \frac{1}{6} b^2 \bar{D}^2 
    	\textbf{W}(\hat{\mathbf{n}}),
\end{equation}
where $S_0=S(b=0)$ is the unweighted signal, $\textbf{D}$ and $\textbf{W}$ are the diffusion and kurtosis tensors, respectively, with $\bar{D}=tr(\textbf{D})$, as defined in \cite{HANSEN2016}, and Einstein's summation convention is implied. 
Let us consider a voxel with fibres oriented according to a Watson ODF.
 Following an analogous procedure as in \cite{NOVIKOV2018}, we can expand the signal $S(b,\hat{\mathbf{n}})$ in Eq. \ref{eq:IntegratedKernel} up to order $\mathcal{O}(b^2)$ according to Eq. \ref{Eq:Cumulant4thOrder_SDE}. This gives a mapping between the biophysical parameter (BP) space and the diffusion kurtosis (DK) space, 
removing the dependence with the acquisition settings and simplifying the analysis of whether different sets of model parameters produce the same signal profile.

Due to the axial symmetry of the Watson distribution, the corresponding diffusion and kurtosis tensors can be expressed in terms of the projection, $\xi = \hat{\mathbf{n}} \cdot \hat{\pmb{\mu}}$, of the gradient direction to the main direction $\hat{\pmb{\mu}}$ \cite{JESPERSEN2017}:
\begin{equation}
\begin{aligned}\label{Eq:BP2DKIWATSONTENSORS}
D(\xi) & = \bigl( f D_\text{a} + (1-f) \Delta_\text{e} \bigr) h_2(\xi,\kappa) + (1-f) D_{\text{e}}^{\perp}, \\
W(\xi) \bar{D}^2 & = 3 \biggl[ \bigl( f D_\text{a}^2 + (1-f) \Delta_{\text{e}}^2 \bigr) h_4(\xi,\kappa) + 2 (1-f) \Delta_{\text{e}} D_{\text{e}}^{\perp} h_2(\xi,\kappa) + (1-f) D_{\text{e}}^{\perp 2} - D(\xi)^2 \biggr],
\end{aligned}
\end{equation}
where $h_2(\xi,\kappa)=\frac{1}{3}+\frac{2}{3} \, p_2 P_2(\xi) \, $ and $h_4(\xi,\kappa)=\frac{1}{5} + \frac{4}{7} p_2 P_2(\xi) + \frac{8}{35} p_4 P_4(\xi)$ are defined as in \cite{JESPERSEN2017}. $P_2(\xi)$ and $P_4(\xi)$ are the second and fourth order Legendre polynomials, and $p_2$, $p_4$ the non-zero second and fourth order coefficients of the spherical harmonics expansion of the Watson distribution:
\begin{equation}\label{Eq:WatsonSHCoeff}
\begin{aligned}
p_2&=\frac{1}{4}\bigg[\frac{3}{\sqrt{\kappa}F(\sqrt{\kappa})}-2-\frac{3}{\kappa}\bigg],\\
p_4&=\frac{1}{32\kappa^2}\bigg[105+12\kappa(5+\kappa)+\frac{5\sqrt{\kappa}(2\kappa-21)}{F(\sqrt{\kappa})}\bigg],
\end{aligned}
\end{equation}
where $F$ denotes the Dawson function \cite{ABRAMOWITZ1972}.
Using these equations, we can derive the relations between the BP and DK parameters that fully describe this axially symmetric environment, as done in \cite{HANSEN2017} for fully aligned fibres, but here for an arbitrary value of $\kappa$:
\begin{equation}\label{Eq:BP2DKIWATSON}
\begin{aligned}
D_\parallel &= 
	\bigl( f D_\text{a} + 
    	(1-f) \Delta_\text{e}\bigr) 
        \,h_2(1,\kappa) + 
    (1-f) D_\text{e}^\perp , \\
D_\perp &= 
	\bigl( f D_\text{a} + 
    	(1-f) \Delta_\text{e}\bigr) 
        \,h_2(0,\kappa) + 
    (1-f) D_\text{e}^\perp , \\
\frac13 W_\parallel \bar{D}^2 + D_\parallel^2 &=
	\bigl( f D_\text{a}^2 + 
    	(1-f) \Delta_{\text{e}}^2 \bigr)
    \,h_4(1,\kappa) + 
    2(1-f) \Delta_\text{e} D_\text{e}^\perp
    	\,h_2(1,\kappa) +
    (1-f) D_\text{e}^{\perp 2}, \\
\frac13 W_\perp \bar{D}^2 + D_\perp^2 &= 
	\bigl( f D_\text{a}^2 + 
    	(1-f) \Delta_{\text{e}}^2 \bigr)
    \,h_4(0,\kappa) + 
    2(1-f) \Delta_\text{e} D_\text{e}^\perp
    	\,h_2(0,\kappa) +
    (1-f) D_\text{e}^{\perp 2}, \\
\frac{5\bar{W} \bar{D}^2}{8} - 
	\frac{ W_\perp \bar{D}^2}{4} - 
    \frac{W_\parallel \bar{D}^2}{24}+
	\frac{(D_\perp + D_\parallel)^2}{4} & = 
	\bigl( f D_\text{a}^2 + 
    	(1-f) \Delta_{\text{e}}^2 \bigr)
    \,h_4\bigl(\tfrac{1}{\sqrt{2}},\kappa\bigr) + 
    2(1-f) \Delta_\text{e} D_\text{e}^\perp
    	\,h_2\bigl(\tfrac1{\sqrt{2}},\kappa\bigr)+
    (1-f) D_\text{e}^{\perp 2},
\end{aligned}
\end{equation}
where $\bar{D} = (2 D_\perp + D_\parallel)/3$. Taking the limit for $\kappa \rightarrow \infty$ we recover the system of equations for parallel fibres presented in \cite{HANSEN2017} (Eq.~12). 

In contrast with the claim in Hansen~{\it et al.}~\cite{HANSEN2017}, even in this extreme case of parallel fibres leaving only four unknowns, these five equations are independent. This is possible due to the nonlinear nature of the system. 
If $\kappa$ is known and not zero (including the limiting case $\kappa\to\infty$ of parallel fibres), the full-system is invertible as long as $f$ is not 0 or 1, and $D_\text{e}^\perp$ is not null. In that case, each point in the DK parameter space (signal profile) corresponds to a single set of BP parameters. However, this is not the case for an arbitrary unknown $\kappa$.
%
%
%
%
Here, the full-system has 5 independent equations with 5 unknowns, but, depending on the parameter values, it can have only one or multiple solutions. This latter case makes the inverse mapping an ill-posed problem.
%

\begin{table}
\caption{Illustration of sets of biophysical (BP) parameter values resulting in the same diffusion--kurtosis (DK) parameters. Each plus or minus branch can correspond to a single, multiple, or none BP parameters. Some sets of BP parameters fall outside the region of plausible parameters, like the $+$ branch solution of the third example. We can observe that the tensor $\textbf{Z}$, incorporated by DDE, discriminates between the BP parameter sets having the same exact DK representation. All diffusivities are in $\mu m^2/ms$ and the $Z$ components in $\mu m^4/ms^2$.}\label{table_Multimodality}
%
%
%
\begin{threeparttable}
\begin{tabular}{*4{c@{,\quad}>{\!\!\!\!}}cc*4{c@{,\quad}>{\!\!\!\!}}ccc}
\headrow
\multicolumn{5}{c}{\bf DK parameters} & \bf Branch & \multicolumn{5}{c}{\bf BP parameters} &\multicolumn{2}{c}{\bf DDE $\mathbf{Z}$ tensor} \\
\headrow
$[D_\parallel$ & $D_\perp$ & $W_\parallel$ & $W_\perp$ & $\bar{W}]$ &  & $[f$ & $D_\text{a}$ & $D_{\text{e}}^{\parallel}$ & $D_{\text{e}}^{\perp}$ & $\kappa]$ & $\zeta_1$ & $\zeta_2$ \\
\hiderowcolors
\rowcolor{white}
$[1.503$ & $0.195$ & $1.456$ & $0.291$ & $0.926]$ & $+$ & $[0.730$ & $2.000$ & $1.000$ & $0.300$ & $8.000]$ & $-0.006$ & $0.210$ \\
\rowcolor{gray!15}
\multicolumn{5}{c}{} & $-$ & $[0.607$ & $1.287$ & $2.191$ & $0.318$ & $11.49]$ &  $0.023$ &  $0.053$ \\
\rowcolor{white}
$[1.557$ & $1.048$ & $0.396$ & $0.708$ & $0.330]$ & $+$ & $[0.250$ & $2.370$ & $1.300$ & $1.390$ & $50.00]$ & $0.349$ & $0.624$ \\
\rowcolor{gray!15}\multicolumn{5}{c}{} & $-$ & \multicolumn{5}{c}{-} & \multicolumn{2}{c}{-} \\
\rowcolor{white}
$[0.457$ & $0.408$ & $2.901$ & $2.702$ & $2.770]$ & $+$ &  $[0.879$ & $1.320$ & $1.401$ & $-0.232$ & $0.265]$ & $-0.190$ & $0.022$ \\
\rowcolor{gray!15}\multicolumn{5}{c}{} & $-$ &    $[0.870$ & $0.950$ & $2.000$ & $0.720$ & $0.360]$ & $-0.023$ & $0.014$ \\
                  \multicolumn{5}{c}{} & $-$ &     $[0.549$ & $0.182$ & $1.071$ & $0.766$ & $1.414]$ & $0.154$ & $-0.002$ \\
                  \multicolumn{5}{c}{} & $-$ &    $[0.510$ & $0.076$ & $0.931$ & $0.794$ & $3.187]$ & $0.161$ & $-0.005$ \\
\rowcolor{white} $[1.560$ & $1.256$ & $0.423$ & $0.540$ & $0.506]$ & $+$ & \multicolumn{5}{c}{-} & \multicolumn{2}{c}{-} \\
\rowcolor{gray!15}\multicolumn{5}{c}{} & $-$ &   $[0.240$ & $1.450$ & $2.100$ & $1.400$ & $2.330]$  & $0.237$  & $0.125$ \\
                  \multicolumn{5}{c}{} & $-$ &    $[0.189$ & $0.668$ & $1.887$ & $1.489$ & $5.442]$   & $0.325$ & $0.057$ \\
\hline
\end{tabular}
\end{threeparttable}
\end{table}

%
In Jespersen~{\it et al.}~\cite{JESPERSEN2017}, the equivalent to the system in Eq.~\ref{Eq:BP2DKIWATSON} is solved reaching two alternative equations for $\kappa$, $\mathcal{F}_\pm(\kappa) =  0$, each giving a branch of solutions. This suggested that, in general, there should be two solutions, one for each branch. However, this is not always the case, as illustrated in Table \ref{table_Multimodality}. 
We derive here an alternative simpler expression of the solution without explicit branches. First, Eq.~\ref{Eq:BP2DKIWATSON} can be reparametrized as:
\begin{equation}\label{Eq:reparametrisation}
\alpha = f D_\text{a} + (1-f) \Delta_\text{e}, \hspace{1em}
\beta  = (1-f) D_\text{e}^\perp, \hspace{1em}
\gamma = f D_\text{a}^2 + (1-f) \Delta_\text{e}^2, \hspace{1em}
\delta = (1-f) \Delta_\text{e} D_\text{e}^\perp, \hspace{1em}
\epsilon = (1-f) D_\text{e}^{\perp 2}.
\end{equation}
If we consider $\kappa$ as known, this gives a linear system of five equations with five unknowns, decoupled into two independent smaller systems:
\begin{equation}\label{Eq:DecoupledMatrices}
\begin{bmatrix}
    D_\parallel \\
    D_\perp
\end{bmatrix}
=
\begin{bmatrix}
    h_2(1,\kappa) & \!\!\!\! 1 \\
    h_2(0,\kappa) & \!\!\!\! 1
\end{bmatrix}
\begin{bmatrix}
    \alpha \\
    \beta
\end{bmatrix}
=
\mathbf{L}
\begin{bmatrix}
    \alpha \\
    \beta
\end{bmatrix}, \hspace{2em}
\begin{bmatrix}
    \frac13 W_\parallel \bar{D}^2 + D_\parallel^2 \\
    \frac13 W_\perp \bar{D}^2 + D_\perp^2 \\
    \frac{5\bar{W} \bar{D}^2}{8} -
    \frac{ W_\perp \bar{D}^2}{4} -
    \frac{W_\parallel \bar{D}^2}{24} +
    \frac{(D_\perp + D_\parallel)^2}{4}
\end{bmatrix}
=
\begin{bmatrix}
    h_4(1,\kappa) & \!\!\!\! 2 h_2(1,\kappa) & \!\!\!\! 1 \\
    h_4(0,\kappa) & \!\!\!\! 2 h_2(0,\kappa) & \!\!\!\! 1 \\
    h_4(\frac{1}{\sqrt{2}},\kappa) & \!\!\!\! 2 h_2(\frac{1}{\sqrt{2}},\kappa) & \!\!\!\! 1
\end{bmatrix}
\begin{bmatrix}
    \gamma \\
    \delta \\
    \epsilon
\end{bmatrix}
=
\mathbf{M}
\begin{bmatrix}
    \gamma \\
    \delta \\
    \epsilon
\end{bmatrix},
\end{equation}
The solution is unique as long as matrices $\mathbf{L}$ and $\mathbf{M}$ are invertible. This is the case when $\kappa \neq 0$, since $\det \mathbf{L}=p_2$ and $\det\mathbf{M}=-\frac12 p_2 p_4$. 
In the limit of a fully isotropic medium ($\kappa=0$) the system has only two independent equations, not allowing the recovering of the kernel parameters without additional information.
By solving the two systems in Eq.~\ref{Eq:DecoupledMatrices} we find expressions for $\alpha$,$\beta$,$\gamma$,$\delta$ and $\epsilon$ that only depend on $\kappa$ and the DK parameters (see Appendix~\ref{AppendixInverseMatrices} for solution). Those variables are actually defined from only 4 kernel parameters (Eq.~\ref{Eq:reparametrisation}), resulting in the coupling equation
\begin{equation}\label{Eq:GammaEquation}
\gamma (\epsilon - \beta^2) = \alpha^2 \epsilon + \delta^2 - 2 \alpha \beta \delta.
\end{equation}
By plugging the expressions for $\alpha$,$\beta$,$\gamma$,$\delta$ and $\epsilon$ as functions of $\kappa$ into Eq.~\ref{Eq:GammaEquation}, we obtain a nonlinear equation for $\kappa$ with potentially multiple solutions. Each solution for $\kappa$ gives a single solution for $\alpha$, $\beta$, $\gamma$, $\delta$ and $\epsilon$, which in turn, gives a single solution for the kernel parameters:
\begin{equation}
f = 1 - \frac{\beta^2}{\epsilon}, \hspace{2em}
D_\text{a} = \frac{\alpha \epsilon - \beta \delta}{\epsilon - \beta^2}, \hspace{2em}
\Delta_\text{e} = \frac{\delta}{\beta}, \hspace{2em}
D_\text{e}^\perp = \frac{\epsilon}{\beta}.
\end{equation}
Thus, the number of solutions to Eq.~\ref{Eq:GammaEquation} corresponds to the number of BP parameter sets that have the same DK parameters. 
Table \ref{table_Multimodality} presents cases with up to 4 solutions. We computed the number of solutions for $10$k random points in the BP parameter space. Most present 2 solutions ($70.2\%$), some only 1 ($29.3\%$), and only a small proportion have 4 solutions ($0.5\%$). 
This gives rise to the previously discussed degeneracy in the model parameter estimation from noisy measurements \cite{JELESCU2015b}. Using very high b-values is an option to solve this problem, as it will add higher order terms in Eq.~\ref{Eq:Cumulant4thOrder_SDE}. However, it is unfeasible in most clinical scanners. Another solution that does not require powerful gradients is to seek for independent measurements providing new information. 
%
%
%
%


\subsection{Model extension to DDE}
DDE adds an extra dimension to the dMRI acquisition, unexplored by SDE experiments. 
For a general multidimensional acquisition \cite{WESTIN2014,WESTIN2016}, due to the Gaussian impermeable compartments, the signal can be written as:
\begin{equation}\label{eq:IntegratedKernelTensor}
 	S_{\text{NODDIDA}}(\mathbf{B})= 
         S_0 \int_{\mathbb{S}^2}\mathcal{P}(\hat{\mathbf{u}})
        \mathcal{K}(\mathbf{B},\hat{\mathbf{u}}) \,
     dS_{\hat{\mathbf{u}}},
 \end{equation}
with the kernel
\begin{equation}\label{eq:Kernel_Btensor}
	\mathcal{K}(\mathbf{B},\hat{\mathbf{u}}) = 
    	f \exp\bigl[- D_{\text{a}}
        	B_{ij} u_i u_j\bigr] +
        (1-f) \exp\bigl[- b D_\text{e}^\perp - 
        	 \Delta_\text{e} 
            B_{ij} u_i u_j\bigr],
\end{equation}
for $b=tr(\mathbf{B})$. The B-tensor of a DDE acquisition is 
$\mathbf{B}=b_1 \, \hat{\mathbf{n}}_{1} \otimes \hat{\mathbf{n}}_{1} + b_2 \, \hat{\mathbf{n}}_{2} \otimes \hat{\mathbf{n}}_{2}$, defined from the pair of gradient directions, $\hat{\mathbf{n}}_1$, $\hat{\mathbf{n}}_2$, and their individual diffusion weightings, $b_1$, $b_2$.  
It has in general two non-zero eigenvalues, \textit{viz.} Planar Tensor Encoding (PTE). In contrast, the SDE's B-tensor, $\mathbf{B}=b\,\hat{\mathbf{n}} \otimes \hat{\mathbf{n}}$, has only one non-zero eigenvalue, \textit{viz.} Linear Tensor Encoding (LTE). Hence, for this model a SDE acquisition is a subset of the DDE acquisitions ($\text{SDE} = \text{DDE}_\parallel \subset \text{DDE}$), for which $\hat{\mathbf{n}}_1=\hat{\mathbf{n}}_2$ (parallel direction pair).  
%
%
%
%
%
%
%
%
%
%
\subsection{DDE information gain}\label{S:DDE_info_gain}
DDE can, in principle, provide independent complementary information. This could transform the inverse mapping of recovering BP parameters from diffusion-weighted measurements into a well-posed problem. 
The fourth order cumulant expansion for the dMRI signal arising from a DDE experiment is
\begin{equation}\label{Eq:CumulantFourthOrder_DDE}
\begin{aligned}
\log (S
/S_0) &= - B_{ij} D_{ij} +\frac12 B_{ij}B_{k\ell} Z_{ijk\ell} \\
&= - (b_1 n_{1i} n_{1j} + b_2 n_{2i} n_{2j} ) D_{ij} + \frac{\bar{D}^2}{6} \, (b_1^2 n_{1i} n_{1j} n_{1k} n_{1\ell} + b_2^2 n_{2i} n_{2j} n_{2k} n_{2\ell}) W_{ijk\ell} + b_1 b_2  \, n_{1i} n_{1j} n_{2k} n_{2\ell} Z_{ijk\ell},
\end{aligned}
\end{equation}
%
%
%
Here, $\textbf{Z}$ is a {\it generalised kurtosis tensor} 
with minor and major symmetries:
\begin{equation}
	Z_{ij\,k\ell}=Z_{ji\,k\ell}=Z_{ij\,\ell k}=Z_{k\ell\,ij},
\end{equation}
but not completely symmetric as $\textbf{W}$, which is obtained from the fully symmetric part of $Z$:
\begin{equation}\label{WSymmeticPartZ}
	\bar{D}^2 W_{ijk\ell}=3Z_{(ijk\ell)}=Z_{ijk\ell}+Z_{i\ell jk}+Z_{ik\ell j}.
\end{equation}
In the case of a Watson ODF, $\textbf{W}$ and $\textbf{Z}$ are transversely isotropic 4th order tensors, \textit{i.e.} they have cylindrical symmetry. Hence, instead of having 15 and 21 independent components they only have 3 and 5, respectively. We can write both tensors as a function of coordinate independent tensor forms, like it is done for $\textbf{W}$ in \cite{HANSEN2017} (Eq.~6):
\begin{equation}\label{WZDecomposed}
	\mathbf{W} = \omega_1 \mathbf{P} + \omega_2 \mathbf{Q} + \omega_3 \mathbf{I} 
    \qquad\text{and}\qquad
    \mathbf{Z} = \frac13 \bar{D}^2 \mathbf{W} + \zeta_1 \mathbf{R} + \zeta_2 \mathbf{J},
\end{equation}
where $\mathbf{Z}$ was written separating its fully symmetric part from the remaining part \cite{ITIN2013}, and
\begin{equation}\label{SymmetricTensors}
\begin{aligned}
  P_{ijk\ell} &= \mu_i \mu_j \mu_k \mu_\ell,\\
  Q_{ijk\ell} &= \frac16 \bigg( \mu_i \mu_j \delta_{k\ell} + \mu_k \mu_\ell \delta_{ij} + \mu_i \mu_k \delta_{j\ell} + \mu_j \mu_k \delta_{i\ell} 
  + \mu_i \mu_\ell \delta_{jk} + \mu_j \mu_\ell \delta_{ik} \bigg),\\
  I_{ijk\ell} &= \frac13 \bigg( \delta_{ij} \delta_{k\ell} + \delta_{ik} \delta_{j\ell} + \delta_{i\ell} \delta_{jk} \bigg),\\
  R_{ijk\ell} &= \frac12 \bigg( \mu_i \mu_j \delta_{k\ell} + \mu_k \mu_\ell \delta_{ij} \bigg) - \frac14 \bigg( \mu_i \mu_k \delta_{j\ell} + \mu_j \mu_k \delta_{i\ell} 
  + \mu_i \mu_\ell \delta_{jk} + \mu_j \mu_\ell \delta_{ik} \bigg),\\
  J_{ijk\ell} &= \delta_{ij} \delta_{k\ell} - \frac12 \bigg( \delta_{ik} \delta_{j\ell} + \delta_{i\ell} \delta_{jk} \bigg),
\end{aligned}
\end{equation}
where $\delta_{ij}$ is the Kronecker delta and $\hat{\pmb{\mu}}$ the Watson distribution main direction. Eq.~\ref{WZDecomposed} shows explicitly that $\textbf{Z}$ contains two extra degrees of freedom independent of $\textbf{W}$. Observe that the fully symmetric part of $\mathbf{R}$ and $\mathbf{J}$ vanishes, so that the information encoded in $\zeta_1$ and $\zeta_2$ is not accessible from a SDE experiment \cite{JESPERSEN2013}. We can isolate the new non-symmetric components by the antisymmetrization
\begin{equation}\label{ZAntisymmetric}
    Z_{ijk\ell} - Z_{ikj\ell}=
    \zeta_1 (R_{ijk\ell} - R_{ikj\ell}) + 
    \zeta_2 (J_{ijk\ell} - J_{ikj\ell}).
\end{equation}
Considering a coordinate frame with the $z$-axis parallel to the fibers main direction $\hat{\pmb{\mu}}$, we can identify
\begin{equation}\label{IsolatedZNewComponents}
    Z_{xxyy} - Z_{xyxy} = \frac32\zeta_2
    \qquad\text{and}\qquad
    Z_{xxzz} - Z_{xzxz} - Z_{xxyy} + Z_{xyxy} = \frac34\zeta_1.
\end{equation}
%
%
%
%
%

Similarly to Eq.~\ref{Eq:BP2DKIWATSON} we can relate the elements of \textbf{Z} to the biophysical parameters like it was done for $\textbf{W}$. For the SM, including NODDIDA, 
the total diffusion and generalised kurtosis tensors are given by the averages
\begin{equation}
\begin{aligned}
D_{ij} & = \bigl\langle D_{ij} \bigr\rangle = \sum_\alpha f_\alpha D_{ij}^{(\alpha)}, \\
Z_{ijk\ell} & = \Bigl\langle \bigl( D_{ij}-\langle D_{ij}\rangle\bigr) \bigl( D_{k\ell}-\langle D_{k\ell}\rangle\bigr)  \Bigr\rangle = \sum_\alpha f_\alpha D_{ij}^{(\alpha)} D_{k\ell}^{(\alpha)} - D_{ij} D_{k\ell},
\end{aligned}
\end{equation}
where $D_{ij}^{(\alpha)}$ denotes the diffusion tensor for each compartment $\alpha$, including in this summation the integral over the unit sphere with the ODF (\textit{cf.} Eq.~\ref{eq:IntegratedKernel}). 
%
%
%
This results in
\begin{equation}
\begin{aligned}
	D_{ij} & =  
    \bigl[ f D_{\text{a}} + (1-f) \Delta_\text{e} \bigr] 
    	H_{ij}^{(2)} +
	(1-f) D_\text{e}^{\perp} \delta_{ij}, \\
	Z_{ijk\ell} & = 
    \bigl[ f D_{\text{a}}^2 + (1-f) \Delta_\text{e}^2 \bigr] 
    	H_{ijk\ell}^{(4)} +
	(1-f) D_\text{e}^{\perp} \Delta_{\text{e}} 
    	\biggl( \delta_{ij} H_{k\ell}^{(2)} +
        	\delta_{k\ell} H_{ij}^{(2)} \biggr) +
	(1-f)  D_\text{e}^{\perp 2} \delta_{ij} \delta_{k\ell} - 
    D_{ij}D_{k\ell},
\end{aligned}
\end{equation}
where
\begin{equation}
H_{ij}^{(2)}  =  \int_{\mathbb{S}^2}  \mathcal{P}(\hat{\mathbf{u}}) \, u_i u_j \, dS_{\hat{\mathbf{u}}} \qquad\text{and}\qquad
H_{ijk\ell}^{(4)}  =  \int_{\mathbb{S}^2}  \mathcal{P}(\hat{\mathbf{u}})\, u_i u_j u_k u_\ell \, dS_{\hat{\mathbf{u}}}.
\end{equation}
For NODDIDA we get $h_2(\xi,\kappa)=H_{ij}^{(2)} n_i n_j$ and $h_4(\xi,\kappa)=H_{ijk\ell}^{(4)} n_i n_j n_k n_\ell$, with $\xi = \hat{\pmb{\mu}} \cdot \hat{n}$. 
The cross-terms of $\mathbf{Z}$ present new information not accessible from SDE. This makes the DDE signal able to resolve the degeneracy. To make this explicit, we can write the components isolated in Eq.~\ref{IsolatedZNewComponents} in the adapted coordinate frame in terms of BP parameters:
%
%
%
%
%
%
%
%
%
%
\begin{equation}
\begin{aligned}
	\frac32\zeta_2&=
    Z_{xxyy} - Z_{xyxy} = 
	(1-f) \biggl[ D_\text{e}^{\perp} \Delta_{\text{e}} 
    	\biggl( H_{xx}^{(2)} + H_{yy}^{(2)} \biggr)
	+ {D_\text{e}^{\perp}}^2 \biggr] 
    - D_{xx}D_{yy} \\
    &=  (1-f) \biggl[ 2 D_\text{e}^{\perp} \Delta_{\text{e}} 
    	h_2(0,\kappa) +
		{D_\text{e}^{\perp}}^2 \biggr] 
	- D_\perp{}^2 \\
    \frac34\zeta_1 &=
    Z_{xxzz} - Z_{xzxz} - Z_{xxyy} + Z_{xyxy} =
    (1-f) D_\text{e}^{\perp} \Delta_{\text{e}} 
    	\Bigl( H_{zz}^{(2)} - H_{yy}^{(2)} \Bigr) 
    - D_{xx}(D_{zz} - D_{yy}) \\
    &= (1-f) D_\text{e}^{\perp} \Delta_{\text{e}} 
    	\bigl( h_2(1,\kappa) - h_2(0,\kappa)  \bigr) 
    - D_\perp(D_\parallel - D_\perp)
\end{aligned}
\end{equation}
Those 2 equations are independent to the ones in Eq.~\ref{Eq:BP2DKIWATSON}, adding complementary information for the mapping between DK and BP spaces
(see last column in Table \ref{table_Multimodality}).
%
%
%
%
Using the same variables defined in Eq.~\ref{Eq:reparametrisation} we get 
\begin{equation}\label{eq:AdditionalDDEequations}
	2h_2(0,\kappa)\delta  + \epsilon = \frac32\zeta_2 + D_\perp{}^2 
    \qquad \text{and}\qquad
    \bigl( h_2(1,\kappa) - h_2(0,\kappa)  \bigr) \delta = \frac34 \zeta_1 + D_\perp (D_\parallel-D_\perp)
\end{equation}These two equations enlarge the system in Eq.~\ref{Eq:DecoupledMatrices}. Following the derivation in Appendix~\ref{app:KappaFromDDE}, we demonstrate that they determine a single solution for $\kappa$:
\begin{equation}\label{eq:KappaFromDDE}
	\frac{h_4(1,\kappa)}{h_4(0,\kappa)}=
    \frac{
    	\frac13 W_\parallel \bar{D}^2 
        - \frac32 ( \zeta_1 + \zeta_2 ) 
        + (D_\parallel - D_\perp)^2 
    }{
    	\frac13 W_\perp \bar{D}^2 - \frac32 \zeta_2
    }
\end{equation}
since the left-hand side is a strictly monotone increasing function on $\kappa$.


\newpage

\section{methods}
\subsection{Signal generation}

All synthetic measurements were generated from substrates composed of $1\mu m$ diameter cylinders to simultaneously assess our stick approximation. We computed the signal attenuation in the cylinder's perpendicular plane with the Gaussian Phase Approximation (GPA) for both SDE \cite{NEUMAN1974} and DDE \cite{IANUS2016}.
%
%
%

Since there is no closed analytical solution for the integral on the sphere in Eq.~\ref{eq:IntegratedKernelTensor},
we computed the spherical convolution using Lebedev's quadrature \cite{LEBEDEV1999}:
\begin{equation}\label{eq:8}
\int_{\mathbb{S}^2} f(\hat{\mathbf{u}}) dS_{\hat{\mathbf{u}}}
\approx  \sum_{i} w_i f(\hat{\mathbf{u}}_i),
\end{equation}
where $w_i$ are the quadrature weights of each grid point $\hat{\mathbf{u}}_i$ across the unit sphere. 
For all configurations of SDE and DDE we used 1,202 quadrature points, which guarantee an exact result up to a $59^{th}$ order spherical harmonics decomposition of the ODF. No practical differences were found between the results from our SDE implementation and the analytic summation for SDE in \cite{ZHANG2011}.

Finally, Rician noise was added to the synthetic signals, normalising it to obtain a $\text{SNR}=50$ for the $b_0$ measurements, like in \cite{JELESCU2015b}.

\subsection{Parameter estimation algorithm}
Parameter estimation was based on maximum likelihood. 
Since, for high SNR, Rician noise can be approximated as Gaussian \cite{GUDBJARTSSON1995}, we used the Trust Region Reflective algorithm implemented in the MATLAB (R2016a, MathWorks, Natick, MA, USA) optimisation toolbox, with the objective cost function
\begin{equation}
F(\boldsymbol{\theta}) = 
\sum_i^N 
\bigl( S(\mathbf{B}_i,\boldsymbol{\theta}) - S_\text{NODDIDA}(\mathbf{B}_i,\boldsymbol{\theta}) \bigr)^2,
\end{equation}
where N is the total number of measurements, $\mathbf{B}_i$ indicates the b-tensor used in the $i$-th measurement and $\boldsymbol{\theta}=[f,D_\text{a}, D_{\text{e}}^{\parallel}, D_{\text{e}}^{\perp}, \kappa]$ is a vector containing the model parameters.
The main direction of the fibres, $\boldsymbol{\hat{\mu}}$, and $S_0$ were fitted independently in a first stage through a DTI fitting like in \cite{JELESCU2015b}. 
For all configurations,
this optimisation procedure was repeated using 30 independent random initialisations for the model parameters to avoid local minima of the five-dimensional cost function. The local solution with the lowest residue was the global optimum.
%
%
%
%
%
%
%
%
%
%
\subsection{SDE and DDE tested configurations}
\begin{figure}[htbp]
\centering
\includegraphics[scale=.21]{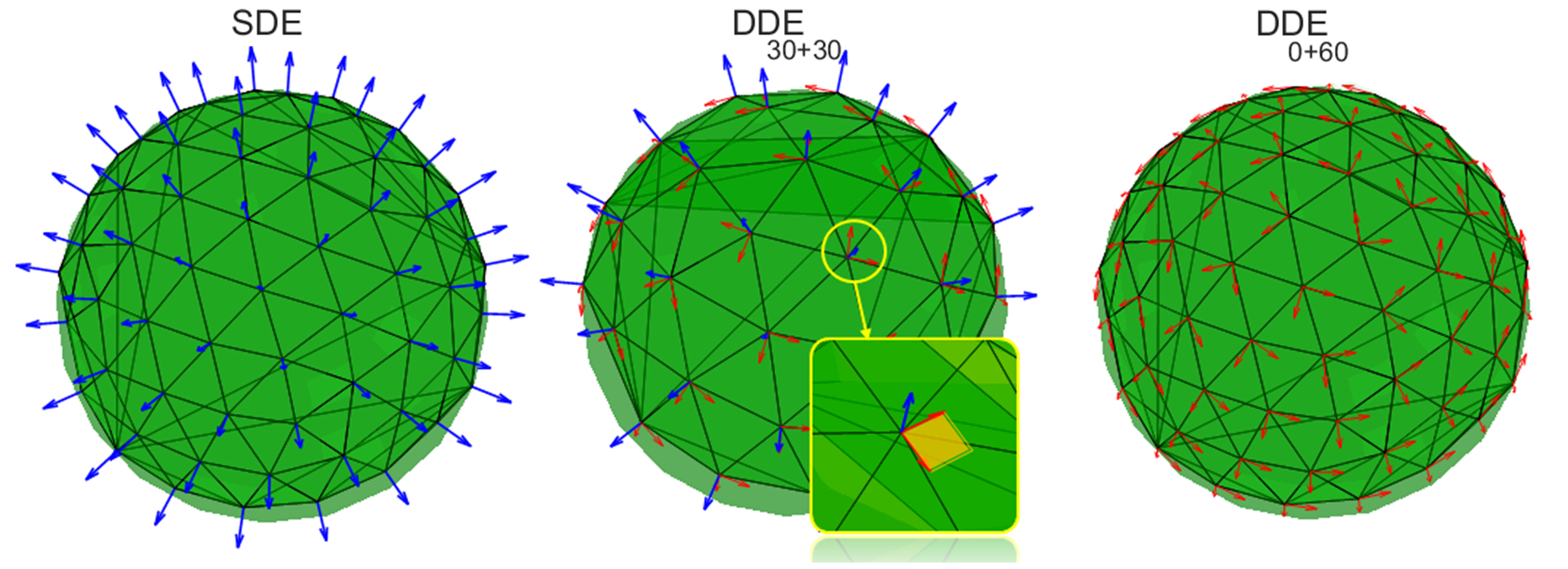}
\caption{Diagram of different measurement protocols (SDE, $\text{DDE}_{30+30}$, and $\text{DDE}_{0+60}$). Only SDE and $\text{DDE}_{30+30}$ were used in experiment 1, while they all were used in experiment 2. Blue colours denote the SDE directions or DDE parallel direction pairs. DDE perpendicular direction pairs are in red.}\label{fig:Experiment_directions}
\end{figure}

Five encoding configurations were considered: $\text{DDE}_{60+0}=\text{SDE}$, $\text{DDE}_{40+20}$, $\text{DDE}_{30+30}$, $\text{DDE}_{20+40}$, and $\text{DDE}_{0+60}$, with progressively increasing proportions of perpendicular direction pairs with respect to parallel direction pairs. We compared the SDE protocol used in \cite{JELESCU2015b} against different DDE acquisitions 
that can be measured in the same experimental time.
%
%
%
%
%
The SDE measurement protocol 
had two shells with b-values of $1$ and $2 \ ms/\mu m^2$ with 30 directions each \cite{JELESCU2015b}. These directions were generated using the Sparse and Optimal Acquisition (SOA) scheme 
\cite{KOAY2012}. DDE configurations were also divided in 2 shells with the same b-values as above and both directions in each pair had equal 
individual diffusion weightings, $b_1=b_2=\frac12 b$. 
Thus, perpendicular direction pairs define axially symmetric planar B-tensors, uniquely defined by their normal vector.
We generated homogeneously distributed normal vectors using the same algorithm used for the SDE directions. The $\text{DDE}_{30+30}$ acquisition had 30 parallel direction pairs and 30 perpendicular direction pairs with normal vectors coinciding with the parallel pairs \cite{COELHO2017} 
(see middle diagram in Fig.~\ref{fig:Experiment_directions}). 
The $\text{DDE}_{0+60}$ protocol had only perpendicular directions pairs
%
%
%
(right diagram in Fig. \ref{fig:Experiment_directions}). 
Configuration $\text{DDE}_{40+20}$ had two parallel per each perpendicular directions pair, and $\text{DDE}_{20+40}$ two perpendicular per each parallel directions pair. All acquisitions had 5 non diffusion-weighted measurements (\textit{i.e.} $b_0$ measurements).
\subsection{Experiments}
%
%
%
%
\begin{wraptable}{r}{6cm}
\vspace{-10pt}
\centering
\begin{tabular}{lcc}
\headrow
\thead{Model parameter} & SET A & SET B \\
$\text{f}$                               & 0.38         & 0.77        \\
$\text{D}_\text{a} \, [\mu m^2/ms]$             & 0.50         & 2.23        \\
$\text{D}_{\text{e}}^{\parallel} \, [\mu m^2/ms]$ & 2.10         & 0.16        \\
$\text{D}_{\text{e}}^{\perp} \, [\mu m^2/ms]$     & 0.74         & 1.48        \\
$\text{c}_2 \, (\kappa)$                  & 0.98 \, (64) & 0.70 \, (4) \\
\hline
\end{tabular}
\caption{Ground truth for experiment 1.}
\label{table_PLIC_GT}
\end{wraptable}
We performed two \textit{in silico} experiments to assess whether the addition of DDE measurements can enhance the parameter estimation in the presence of typical noise in the measurements. 

In the first experiment, we considered two possible instances of NODDIDA parameter values for a voxel in the posterior limb of the internal capsule (PLIC) taken from \cite{JELESCU2015b} (see Table \ref{table_PLIC_GT}), 
for which SDE estimates showed a bimodal distribution. We explored in detail whether DDE solve the degeneracy in these particular cases.
Only SDE and $\text{DDE}_{30+30}$ acquisition configurations were considered for this experiment. Two thousand and five hundred independent realisations of Rician noise were added to the synthetic SDE and DDE signals.

The second experiment aims to compare the accuracy provided by SDE and the different DDE configurations
extensively along the feasible region of the full five-dimensional (5D) space of parameters. 
This allows exploring whether there are subregions presenting different behaviours. 
A 5D grid was generated by all the combinations of $f = 0.1:0.2:0.9$, $D_\text{a}=0.3:0.5:2.3$, $D_{\text{e}}^{\parallel}=0.8:0.5:1.8$, $D_{\text{e}}^{\perp}=0.5:0.5:1.5$, and 
$\kappa=[0.84,2.58,4.75,9.27,15.53,33.70]$. The fraction and the diffusivities were selected from a uniform discretisation of the expected range, and $\kappa$ values were chosen such that the mean-squared-cosine corresponding angle,
$\cos^2\varphi=c_2= 
	\langle (\hat{\mathbf{u}}\cdot \hat{\pmb{\mu}})^2 \rangle = 
    \bigl(2\sqrt{\kappa} F(\sqrt{\kappa})
    	\bigr)^{-1}  
    -(2\kappa)^{-1}$,
was $\varphi=[50^\circ, 40^\circ, 30^\circ, 20^\circ, 15^\circ, 10^\circ]$ ($c_2=[0.41,0.59,0.75,0.88,0.93,0.97]$). We generated 50 independent Rician noise realisations (SNR=50) for the measurements of each combination of the parameters for the five configurations.






\section{results}
\begin{figure}[htbp]
\centering
\includegraphics[scale=.30]{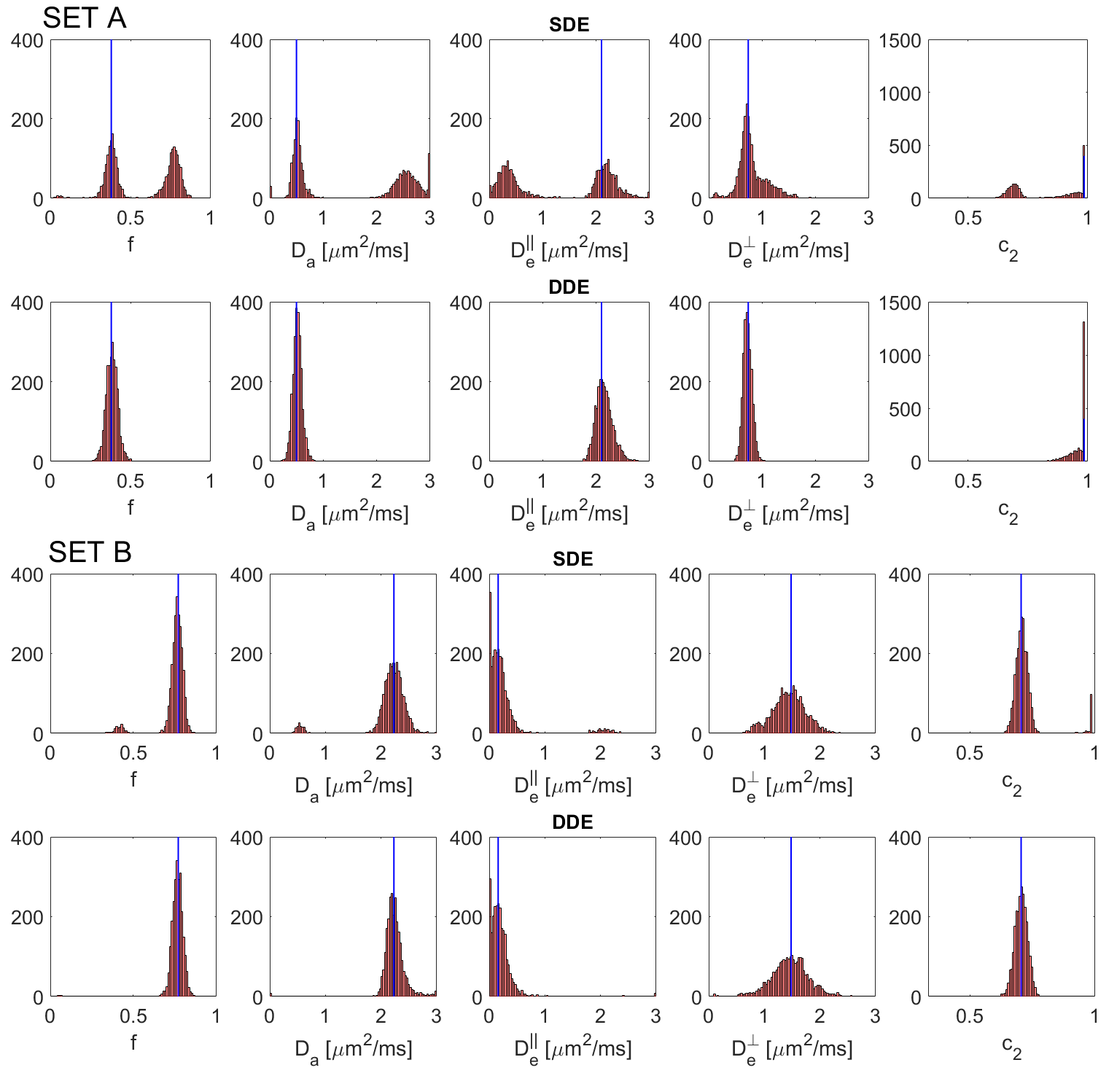}
\caption{Histograms of the estimated model parameters for SDE (top row) and $\text{DDE}_2$ (bottom row) schemes in the first experiment for 2,000 independent
%
%
noise realisations (SNR=50). The ground truth represents two possible solutions of the NODDIDA model applied to a voxel in the PLIC (Table \ref{table_PLIC_GT}). These values are shown in red lines and correspond to set A (upper two rows), and set B (lower two rows).}\label{fig:Exp1_Histograms}
\end{figure}

Histograms of the estimated model parameters from the first experiment 
(Fig.~\ref{fig:Exp1_Histograms}) 
show an increase in the accuracy and stability of the estimates with the DDE scheme.
The bimodal distribution of the estimated parameters is evident with the SDE acquisition, 
confirming that it is not possible to differentiate true and spurious minima. This effect is removed when using the DDE sequence.
%
%
%
%
%

\begin{figure}[htbp]
\centering
\includegraphics[scale=.23]{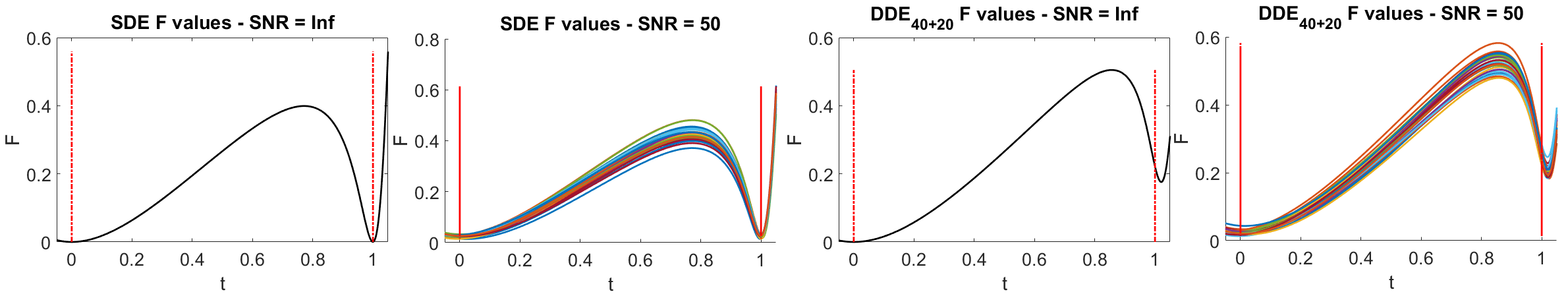}
\caption{Plots of $\text{F}_{\text{A}}(\boldsymbol{\theta}(t))$ for different values of $t \in [-0.05,1.05]$, with  $\boldsymbol{\theta}(t)=t \, \boldsymbol{\theta}_{\text{spur}} + (1-t) \, \boldsymbol{\theta}_{\text{true}}$. Black curves show $\text{F}_{\text{A}}$ values for  noise-free SDE and $\text{DDE}_2$ acquisitions. The coloured curves show 30 independent realisations of $\text{F}_{\text{A}}$ for SNR=50.}\label{fig:1D_cut_F_SET_A}
\end{figure}

We analysed the shapes of the SDE and DDE objective functions from the synthetic measurements of SET A (sum of squared differences: $\text{F}_{\text{A}}(\boldsymbol{\theta})$).
%
%
%
%
%
%
To facilitate the visualisation of these 5D functions, we performed a 1D cut through a straight line joining the true and spurious minima of SDE. This was parametrised with the scalar variable t: $\boldsymbol{\theta}= t \, \boldsymbol{\theta}_{\text{spur}} + (1-t) \, \boldsymbol{\theta}_{\text{true}} \, ; \, t \in [0,1]$, where $\boldsymbol{\theta}_{\text{true}}=[0.38,0.5,2.1,0.74,64]$ and $\boldsymbol{\theta}_{\text{spur}}=[0.78,2.67,0.32,0.85,3.65]$, with diffusivities expressed in $\mu\text{m}^2/\text{ms}$. Figure \ref{fig:1D_cut_F_SET_A} shows the behaviour of $\text{F}_{\text{A}}(\boldsymbol{\theta})$ along this cut as a function of $t$. It can be observed that although the DDE objective function is still bimodal, the spurious and true minima have significantly different absolute values (due to the contribution of the tensor $\mathbf{Z}$ to the DDE signal). This enables us to distinguish both peaks in conditions where SDE cannot (\textit{i.e.} $\text{b}_{\text{max}}=2 ms/\mu m^2$). Adding more directions to the SDE acquisition would not help to differentiate the peaks, as even in the noiseless case these two sets produce the same signal. Only by increasing the SDE diffusion weighting the spurious minimum could be differentiated from the true one.

\begin{figure}[htbp]
\centering
\includegraphics[scale=.205]{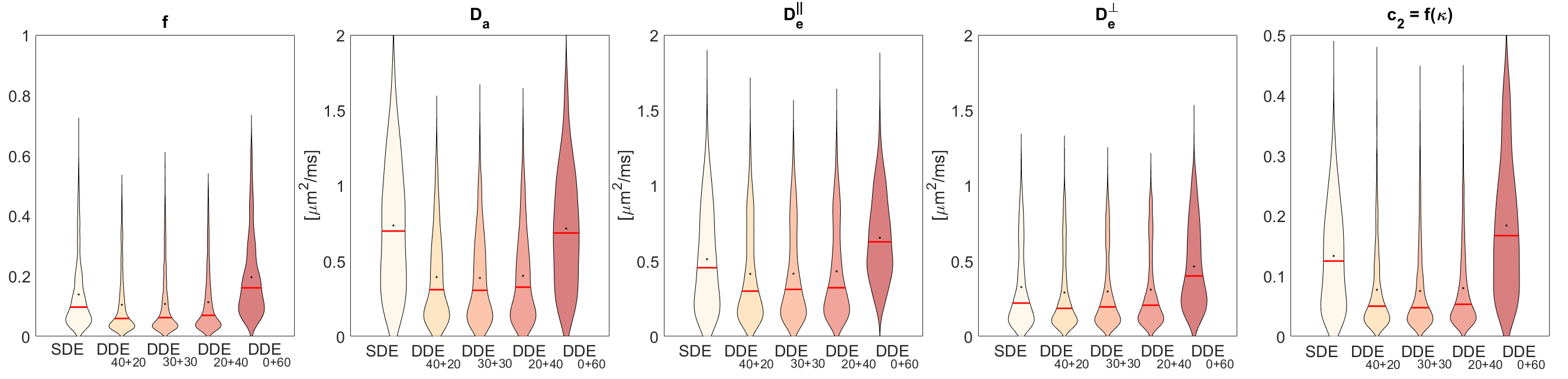}
\caption{Violin plots of the RMSE for all model parameters for all voxels in the 5D grid (a total of $5\times5\times3\times3\times6 = 1,350$). Black dots denote the mean and red lines the median. The RMSE for each voxel was computed by repeating the estimation on 50 independent noise realisations of the measurements for each voxel.}\label{fig:RMSE_violin}
\end{figure}

The Root Mean Square Error (RMSE) of the parameter estimates from the second experiment are summarised in Table \ref{table_MSE_5D_summary} and shown in Fig. \ref{fig:RMSE_violin} with violin plots (similar to box plots but showing a more detailed probability density). For each point in the 5D grid, the RMSE was computed
%
%
%
%
%
%
from 50 independent noise realisations. 
On average, $\text{DDE}_{40+20}$ and $\text{DDE}_{30+30}$ are the most accurate configurations for estimating all parameters. 
This suggests that the incorporation of even a small proportion of DDE measurements can remove the degeneracy, leading to an increase in accuracy and precision.
%
%
%

\begin{table}[htbp]
\caption{Mean and standard deviation of the RMSE over the whole grid for each acquisition protocol and each of the estimated parameters.}
\begin{tabular}{lccccc}
\headrow
\thead{RMSE $(\mu ; \sigma)$} & $f$ & $D_\text{a} [\mu m^2/ms]$ & $D_{\text{e}}^{\parallel} [\mu m^2/ms]$ & $D_{\text{e}}^{\perp} [\mu m^2/ms]$ & $c_2 = f(\kappa)$ \\
SDE            & $0.14$ ; $0.12$ & $0.74$ ; $0.43$ & $0.51$ ; $0.33$ & $0.33$ ; $0.27$ & $0.13$ ; $0.08$ \\
$\text{DDE}_{40+20}$ & $0.10$ ; $0.10$ & $0.39$ ; $0.30$ & $0.41$ ; $0.31$ & $0.29$ ; $0.25$ & $0.08$ ; $0.07$ \\
%
%
$\text{DDE}_{30+30}$ & $0.11$ ; $0.10$ & $0.39$ ; $0.29$ & $0.42$ ; $0.30$ & $0.30$ ; $0.25$ & $0.08$ ; $0.07$ \\
$\text{DDE}_{20+40}$ & $0.11$ ; $0.10$ & $0.40$ ; $0.30$ & $0.43$ ; $0.31$ & $0.31$ ; $0.25$ & $0.08$ ; $0.07$ \\
$\text{DDE}_{0+60}$ & $0.20$ ; $0.13$ & $0.72$ ; $0.38$ & $0.65$ ; $0.28$ & $0.46$ ; $0.27$ & $0.18$ ; $0.11$ \\
\hline  
\end{tabular}\label{table_MSE_5D_summary}
\end{table}

\begin{figure}[htbp]
\centering
\includegraphics[scale=.31]{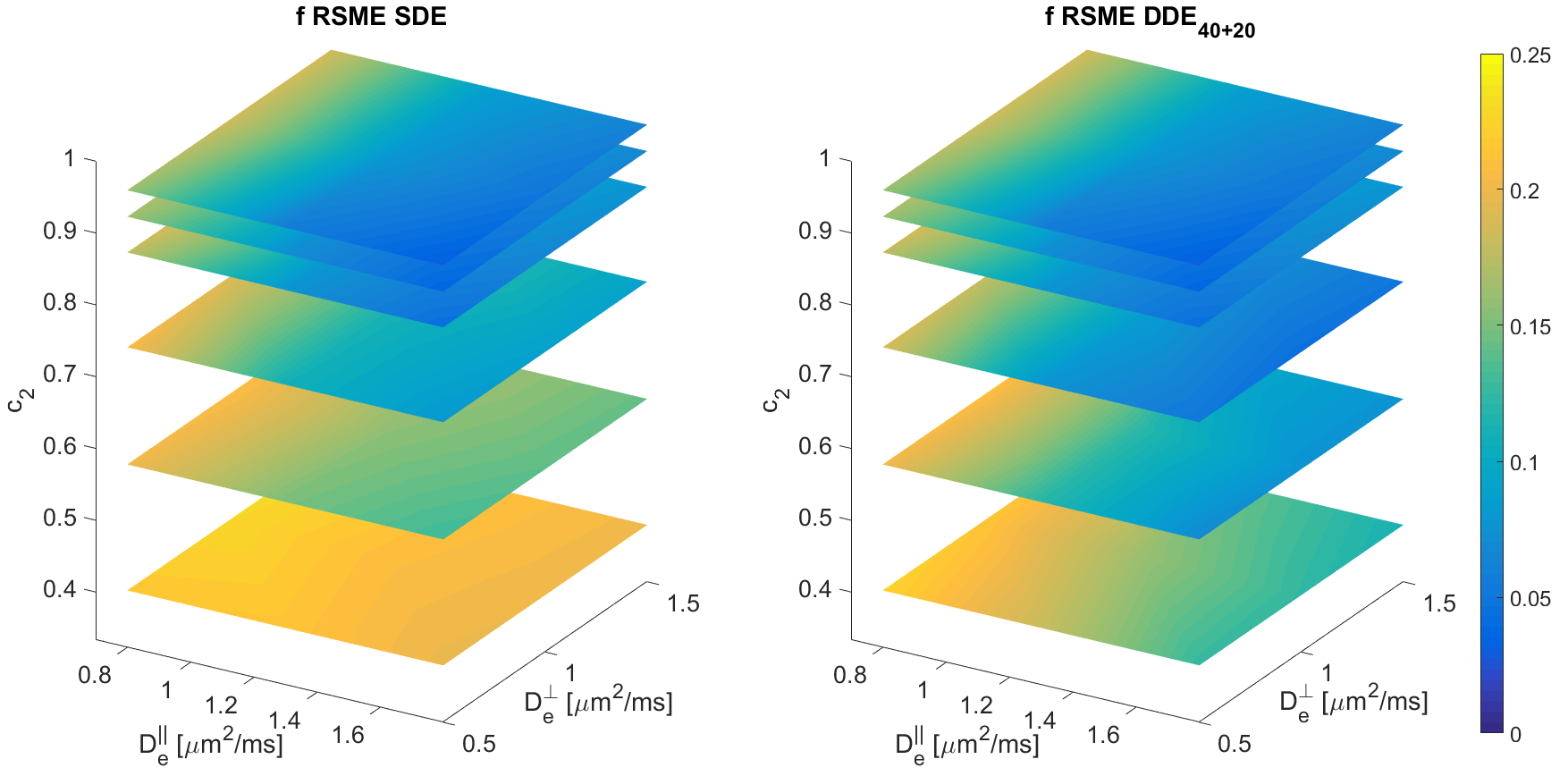}
\caption{RMSE of $f$, for SDE and $\text{DDE}_{40+20}$ acquisition protocols. This 3D plot shows the projection over $D_\text{e}^{\parallel}$, $D_\text{e}^{\perp}$, and $c_2$ of all the RMSE in the 5D grid. This projection was made by computing the square root of the quadratic mean of the errors in the remaining 2 dimensions ($\text{E}_{proj,ijk}=\sqrt{\sum_\ell \sum_m E_{ijk\ell m}^2/(N_\ell N_m)}$).}\label{fig:f_RMSE_3D_f_Di_kappa}
\end{figure}

\begin{figure}[htbp]
\centering
\includegraphics[scale=.31]{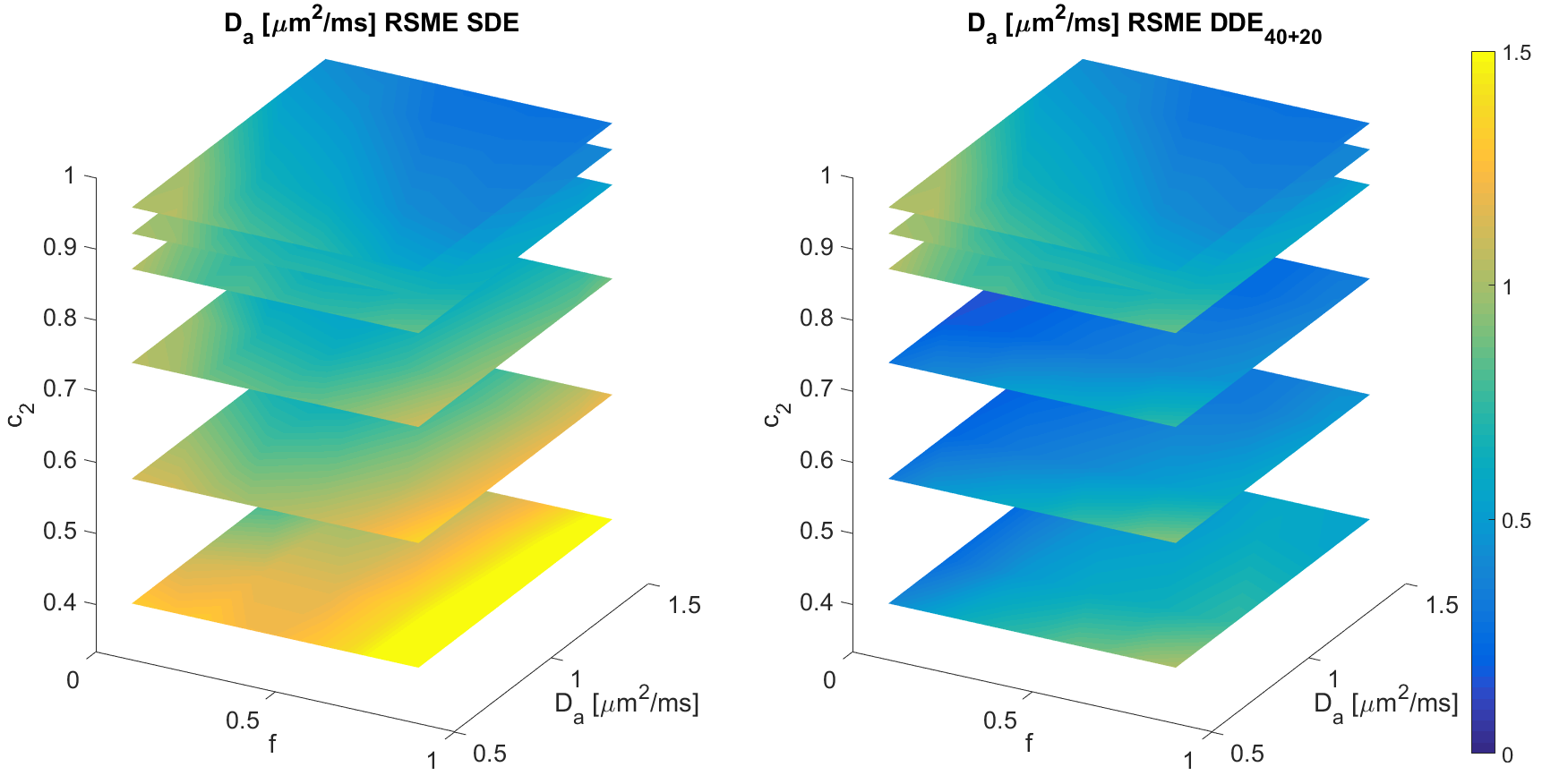}
\caption{RMSE of $D_\text{a}$, for SDE and $\text{DDE}_{40+20}$ acquisition protocols. This 3D plot shows the projection over $f$, $D_\text{a}$, and $c_2$ of all the RMSE in the 5D grid. This projection was made by computing the square root of the quadratic mean of the errors in the remaining 2 dimensions ($\text{E}_{proj,ijk}=\sqrt{\sum_\ell \sum_m E_{ijk\ell m}^2/(N_\ell N_m)}$).}\label{fig:Di_RMSE_3D_kappa_Depar_Deperp}
\end{figure}

To compare the performance of SDE and DDE 
in different regions of the parameter space, we projected the 5D RMSE map onto different 3D sub-spaces. Figures \ref{fig:f_RMSE_3D_f_Di_kappa} and \ref{fig:Di_RMSE_3D_kappa_Depar_Deperp} show two different 3D projections, over $(\text{D}_{\text{e}}^{\parallel}, \text{D}_{\text{e}}^{\perp},c_2(\kappa))$ and over $(f, \text{D}_\text{a},c_2(\kappa))$, 
of the RMSE of $f$ and $D_\text{a}$, respectively.


\newpage

\section{Discussion}


Our work shows that modifying the diffusion MRI pulse sequence can mitigate the degeneracy on NODDIDA’s parameter estimation. Our proposal circumvents the need of presetting diffusivities to \textit{a priori} values as in NODDI.  
%
We showed that estimating the NODDIDA model through SDE is in many scenarios an ill-posed problem. Depending on the specific combination of model parameters, multiple parameter sets may produce the same signal profile. 
We illustrate for a particular voxel the intuition behind the improvements of the proposed methodology. \textit{In silico} experiments over a wide range of model parameter combinations showed that extending the acquisition to DDE makes the inverse problem well-posed and solves the degeneracy in the parameter estimation. Combining DDE parallel (\textit{i.e.} LTE) and perpendicular (\textit{i.e.} PTE) direction pairs not only provides more stable parameter estimates but also increases the precision and accuracy of the parameter estimates. 

In Section \ref{S:THEORY} we showed that in the case of parallel fibres, the inverse problem of recovering biophysical parameters from noiseless SDE measurements is well-posed, but that this is not the case for fibres following a Watson ODF with arbitrary unknown concentration $\kappa$. We demonstrated the existence of multiple BP parameter sets that describe the signal equally well up to intermediate b-values. In contrast, we showed analitically that the $\mathbf{Z}$ tensor includes non-symmetric independent components that are accessible by DDE, but not by SDE, allowing the complete inverse mapping between the DK and BP parameter spaces. 
Consistently, the first experiment showed that in both of the PLIC synthetic voxels, DDE leads to more accurate parameter estimations. This is clearly seen when analysing the optimisation cost-function which shows that although DDE also presents multiple local minima, the global minimum is substantially deeper, unlike SDE, thus it can be picked in typical noise levels. However, two points in the 5D model parameter space are insufficient to draw more general conclusions. Therefore, the second experiment swept the parameter space extensively using a regular grid. Mean results (see Table \ref{table_MSE_5D_summary}) showed 
the highest accuracy for an acquisition consisting of both linear and planar B-tensors, suggesting that the optimal combination for the scenario considered is between $\text{DDE}_{40+20}$ and $\text{DDE}_{30+30}$ configurations.

Increasing the total number of measurements and SNR will have a larger impact in enhancing DDE parameter estimation than with SDE, since the bimodality present in SDE implies a non-zero lower bound for the achievable MSE even without noise. 
Novikov \textit{et al.} \cite{NOVIKOV2018} showed that with a SDE acquisition sensitive up to $\mathcal{O}(b^3)$ the ODF coefficients of the SM can be accurately estimated through the LEMONADE approach. However, for acquisitions sensitive up to $\mathcal{O}(b^2)$ we have shown that even in the case of constraining the shape of the ODF to be a Watson distribution the degeneracy is still present due to the multiple solutions to Eq.~\ref{Eq:GammaEquation}. Results from \cite{FIEREMANS2018} show that the addition of STE data also leads to an increase in the precision of $D_\text{a}$ and $f$ in \textit{in vivo} experiments. In our synthetic experiments the addition of PTE data increases the accuracy in all the parameter estimates (to a lesser extent in $f$ and $D_{\text{e}}^{\perp}$). Recently, Dhital \textit{et al.} \cite{DHITAL2018} showed through \textit{in silico} experiments that incorporating PTE data to LTE data enabled us to discriminate spurious solutions in the cost-function. 
This latter result is explained by our theoretical analysis in Section \ref{S:DDE_info_gain} where we derive the independent equations provided by DDE that make the inverse problem well-posed.
While finalizing this paper, a preprint \cite{REISERT2018} appeared, reaching similar conclusions.

Biophysical models are promising for extracting microstructure-specific information but care must be taken when applying them in dMRI. Some assumptions are more meaningful than others and hence their impact on parameter estimation must be assessed 
\cite{NOVIKOV2018b}. Releasing the diffusivities in the typical two-compartment model eliminates an invalid assumption, reduces possible biases in the estimated parameters, and provides extra information amenable to be used as a biomarker of microstructural integrity and sensitive to specific disease processes \cite{FIEREMANS2012,JELESCU2015a,JELESCU2016}. In this work, we have focused on analysing the estimability of the model under different acquisition settings. The validation against complementary real data is an independent problem. Both should be addressed further to bring biophysical models to the clinic.

Recent work by Novikov \textit{et al.} \cite{NOVIKOV2018} studied the unconstrained SM and concluded that if high b-values are not feasible then orthogonal measurements might be an alternative to uniquely relate the kernel parameters with the signal. 
Veraart \textit{et al.} extended the SM to acquisitions with varying echo time (TE) \cite{VERAART2017}. 
%
%
This latter work goes in a similar direction to our work here, \textit{i.e.} adding extra dimensions to the experiment and changing the objective function to avoid ill-posedness. However, measuring multiple directions while varying the TE implies increasing the acquisition time and 
TE, 
affecting the SNR. However, this approach can be combined with DDE leading to a DDE acquisition with multiple TEs. Recently, Lampinen \textit{et al.} \cite{LAMPINEN2017a} showed that by acquiring data with linear, planar, and spherical tensor encodings the accuracy in estimating the microstructural anisotropy was increased compared to that derived from NODDI's parameters. Additionally, Dhital \textit{et al.} \cite{DHITAL2017} measured the intracellular diffusivity using isotropic encoding. These two works point in a similar direction than ours, \textit{i.e.} extending the acquisition to combine different shapes of b-tensors to maximise the accuracy. Future work will study the generalisation of the model to a multidimensional acquisition. Also, a detailed analysis of the impact of noise will be performed, further assessing the practical identifiability of the model parameters. 

This work's aim was to demonstrate that it is possible to solve the intrinsic degeneracy of the NODDIDA model by using DDE. 
Work by Tariq \textit{et al.} has extended the initial NODDI model to a Bingham ODF \cite{TARIQ2016}. Additionally, Novikov \textit{et al.} 
\cite{NOVIKOV2018} proposed the unconstrained SM with ODF to be described by a series of spherical harmonics.
We plan to extend the analysis in this paper to general ODFs.
The extension of biophysical models to multidimensional dMRI acquisitions \cite{WESTIN2016} should be further explored.
The comparisons made in this work between SDE and DDE protocols do not consider the optimisation of the diffusion directions in DDE, just taking four arbitrary chosen DDE protocols extrapolated from an optimised SDE. 
We expect that further optimisation of the DDE acquisition protocol may also lead to larger improvements. Finally, the largest errors in the parameter estimates occur for $\kappa\rightarrow 0$. This might mean that for highly dispersed tissue (\textit{i.e.} grey matter) many measurements might be needed to accurately estimate model parameters.
%
%


\section{Conclusions}


The potential increase in sensitivity and specificity in detecting brain microstructural changes is a major driving force for developing biophysical models. However, non-linear parameter estimation of these models is not necessarily well-posed and can lead to unreliable parameter values. In this work we not only extended the NODDIDA biophysical model from SDE to DDE schemes, but also demonstrated theoretically the advantages this latter approach has. We illustrated how DDE resolves the degeneracy issue intrinsic to the model estimation from SDE. We prove theoretically that DDE provides complementary information that makes the parameter estimation well-posed. Additionally, this extension leads to an increase in the accuracy and precision in the model parameter estimates in the presence of noise. The combination of parallel and perpendicular measurements for optimal parameter estimation as function of SNR and measurement time remains to be investigated. Our approach does not require high diffusion weightings to make the inverse problem well-posed and it can be further developed for the unconstrained SM.


\section*{acknowledgements}

This work has been supported by the OCEAN project (EP/M006328/1) and MedIAN Network (EP/N026993/1) both funded by the Engineering and Physical Sciences Research Council (EPSRC) and the European Commission FP7 Project VPH-DARE@IT (FP7-ICT-2011-9-601055). DKJ is supported by a Wellcome Trust Investigator Award (096646/Z/11/Z) and a Wellcome Trust Strategic Award (104943/Z/14/Z).

\section*{conflict of interest}
The authors declare no conflict of interest.






\newpage

\appendix
\section{Inverting the full-system for the Watson case}\label{AppendixInverseMatrices}
The system in Eq.~\ref{Eq:DecoupledMatrices} has a unique solution as long as $\kappa \neq 0$ ($\det \mathbf{L}=p_2$ and $\det\mathbf{M}=- \frac12 p_2 p_4$). Their inverse matrices are:
\begin{equation}\label{Eq:inverseMatrices}
\begin{aligned}
\mathbf{L}^{-1}
&=\frac{1}{p_2}
\begin{bmatrix}
    1 & -1 \\
    -\frac13 + \frac13 p_2 & \frac13 + \frac23 p_2
\end{bmatrix}, \\
\mathbf{M}^{-1}
&=\frac{-4}{p_2 p_4}
\begin{bmatrix}
    -p_2/2 & -p_2/2 & p_2 \\
    \frac{3}{14}p_2 - \frac{5}{56}p_4 &
    \frac{3}{14}p_2 + \frac{9}{56}p_4 &
    -\frac{3}{7}p_2 - \frac{1}{14}p_4 \\
    -\frac{3}{70}p_2 + \frac{5}{84}p_4 - \frac{1}{60}p_2 p_4 &
    -\frac{3}{70}p_2 - \frac{3}{28}p_4 - \frac{1}{10}p_2 p_4 &
    \frac{3}{35}p_2  + \frac{1}{21}p_4 - \frac{2}{15}p_2 p_4
\end{bmatrix}.
\end{aligned}
\end{equation}
These provide expressions for $\alpha$,$\beta$,$\gamma$,$\delta$ and $\epsilon$ that only depend on $\kappa$ and the DK parameters.

\section{Single solution for $\kappa$ from DDE}\label{app:KappaFromDDE}
From the systems in Eq.~\ref{Eq:DecoupledMatrices} and Eq.~\ref{eq:AdditionalDDEequations}, we can select 4 equations generating the joint system:
\begin{equation}
\begin{bmatrix}
    h_4(1,\kappa) & 2 h_2(1,\kappa) & 1 \\
    h_4(0,\kappa) & 2 h_2(0,\kappa) & 1 \\
    0 & h_2(1,\kappa)-h_2(0,\kappa)& 0 \\
    0 & 2 h_2(0,\kappa) & 1
\end{bmatrix}
\begin{bmatrix}
    \gamma \\
    \delta \\
    \epsilon
\end{bmatrix}
	=
\begin{bmatrix}
    \frac13 W_\parallel \bar{D}^2 + D_\parallel^2 \\
    \frac13 W_\perp \bar{D}^2 + D_\perp^2 \\
    \frac34 \zeta_1 + D_\perp D_\parallel - D_\perp^2\\
    \frac32 \zeta_2 + D_\perp^2
\end{bmatrix}
\end{equation}
By simple linear combinations we reach
\begin{equation}
\left[ \hspace{-1.1ex}
\begin{array}{ccc|c}
    h_4(1,\kappa) & 2 h_2(1,\kappa) & 1
    	& \frac13 W_\parallel \bar{D}^2 + D_\parallel^2 \\
    h_4(0,\kappa) & 2 h_2(0,\kappa) & 1
    	& \frac13 W_\perp \bar{D}^2 + D_\perp^2 \\
    0 & 2 h_2(1,\kappa)& 1
    	& \frac32 ( \zeta_1 + \zeta_2 ) - D_\perp^2 + 2D_\perp D_\parallel  \\
    0 & 2 h_2(0,\kappa) & 1
    	& \frac32 \zeta_2 + D_\perp^2
\end{array}
\hspace{-1.1ex} \right]
\sim
\left[ \hspace{-1.1ex}
\begin{array}{c@{\ }cc|c}
    h_4(1,\kappa) & 0 & 0
    	& \frac13 W_\parallel \bar{D}^2 - \frac32 ( \zeta_1 + \zeta_2 ) + (D_\parallel - D_\perp)^2 \\
    h_4(0,\kappa) & 0 & 0
    	& \frac13 W_\perp \bar{D}^2 - \frac32 \zeta_2 \\
    0 & 2 h_2(1,\kappa)& 1
    	& \frac32 ( \zeta_1 + \zeta_2 ) - D_\perp^2 + 2D_\perp D_\parallel  \\
    0 & 2 h_2(0,\kappa) & 1
    	& \frac32 \zeta_2 + D_\perp^2
\end{array}
\hspace{-1.1ex} \right]
\end{equation}
Finally, dividing the first and second equations, the dependency on $\gamma$ cancels out, resulting in Eq.~\ref{eq:KappaFromDDE}, providing a single solution for $\kappa$. This is possible since $\gamma$ is strictly positive, unless there is no axons ($f=0$) and the extracellular diffusion is isotropic ($\Delta_\text{e} = 0$), and $h_4(0,\kappa)>0$ for all finite $\kappa$.




\bibliography{Coelho_bibliography}

\begin{thebibliography}{56}
\providecommand{\natexlab}[1]{#1}
\providecommand{\url}[1]{\texttt{#1}}
\providecommand{\urlprefix}{}

\bibitem[{Callaghan(2010)Paul T. Callaghan}]{CALLAGHAN2010}
Callaghan PT.
\newblock Physics of Diffusion.
\newblock In: Jones DK, editor. Diffusion {MRI}: Theory, Methods and
  Applications Oxford: Oxford University Press; 2010.p. 45--56.

\bibitem[{Kiselev(2017)Valerij G. Kiselev}]{KISELEV2016}
Kiselev VG.
\newblock Fundamentals of diffusion {MRI} physics.
\newblock {NMR} in Biomedicine 2017;30:1--18.

\bibitem[{Assaf(2008)Yaniv Assaf}]{ASSAF2008b}
Assaf Y.
\newblock Can we use diffusion {MRI} as a bio-marker of neurodegenerative
  processes?
\newblock BioEssays 2008;30(11-12):1235--1245.

\bibitem[{Basser et~al.(1994)Peter J. Basser and James Mattiello and Denis
  LeBihan}]{BASSER1994}
Basser PJ, Mattiello J, LeBihan D.
\newblock Estimation of the Effective Self-Diffusion Tensor from the {NMR} Spin
  Echo.
\newblock Journal of Magnetic Resonance 1994;103:247--254.

\bibitem[{Assaf and Cohen(1999)Yaniv Assaf and Yoram Cohen}]{ASSAF1999}
Assaf Y, Cohen Y.
\newblock Structural information in neuronal tissue as revealed by q-space
  diffusion {NMR} spectroscopy of metabolites in bovine optic nerve.
\newblock NMR in Biomedicine 1999;12:335--344.

\bibitem[{Tuch(2004)David S. Tuch}]{TUCH2004}
Tuch DS.
\newblock Q-{Ball} {Imaging}.
\newblock Magnetic Resonance in Medicine 2004;52:1358--1372.

\bibitem[{Tournier et~al.(2004)J. Donald Tournier and Fernando Calamante and
  David G. Gadian and Alan Connelly}]{TOURNIER2004}
Tournier JD, Calamante F, Gadian DG, Connelly A.
\newblock Direct estimation of the fiber orientation density function from
  diffusion-weighted {MRI} data using spherical deconvolution.
\newblock NeuroImage 2004;23:1176--1185.

\bibitem[{Jensen et~al.(2005)Jens H. Jensen and Joseph A. Helpern and Anita
  Ramani and Hanzhang Lu and Kyle Kaczynski}]{JENSEN2005}
Jensen JH, Helpern JA, Ramani A, Lu H, Kaczynski K.
\newblock Diffusional {Kurtosis} {Imaging}: {The} Quantification of
  Non-Gaussian Water Diffusion by Means of Magnetic Resonance Imaging.
\newblock Magnetic Resonance in Medicine 2005;53:1432--1440.

\bibitem[{Novikov et~al.(2018)Dmitry S. Novikov and Valerij G. Kiselev and Sune
  N. Jespersen}]{NOVIKOV2018b}
Novikov DS, Kiselev VG, Jespersen SN.
\newblock On modeling.
\newblock Magnetic Resonance in Medicine 2018;79:3172 -- 3193.

\bibitem[{Gelderen et~al.(1994)P. Van Gelderen and D. Despres and P. C. M. Van
  Zijl and C. T. W. Moonen}]{VANGELDEREN1994}
Gelderen PV, Despres D, Zijl PCMV, Moonen CTW.
\newblock Evaluation of restricted diffusion in cylinders phosphocreatine in
  rabbit leg muscle.
\newblock Journal of Magnetic Resonance, Series B 1994;103:255--260.

\bibitem[{Stanisz et~al.(1997)Greg J. Stanisz and Aaron Szafer and Graham A.
  Wright and R. Mark Henkelman}]{STANISZ1997}
Stanisz GJ, Szafer A, Wright GA, Henkelman RM.
\newblock An analytical model of restricted diffusion in bovine optic nerve.
\newblock Magnetic Resonance in Medicine 1997;37:103--111.

\bibitem[{Assaf et~al.(2004)Yaniv Assaf and Tamar Blumenfeld-Katzir and Yossi
  Yovel and Peter J. Basser}]{ASSAF2004a}
Assaf Y, Blumenfeld-Katzir T, Yovel Y, Basser PJ.
\newblock New Modeling and Experimental Framework to Characterize Hindered and
  Restricted Water Diffusion in Brain White Matter.
\newblock Magnetic Resonance in Medicine 2004;52:965--978.

\bibitem[{Jespersen et~al.(2007)Sune N. Jespersen and Christopher D. Kroenke
  and Leif {{\O}}stergaard and Joseph J. H. Ackerman and Dmitriy A.
  Yablonskiy}]{JESPERSEN2007}
Jespersen SN, Kroenke CD, {{\O}}stergaard L, Ackerman JJH, Yablonskiy DA.
\newblock Modeling dendrite density from magnetic resonance diffusion
  measurements.
\newblock NeuroImage 2007;34:1473--1486.

\bibitem[{Zhang et~al.(2012)Hui Zhang and Torben Schneider and Claudia A.
  Wheeler-Kingshott and Daniel C. Alexander}]{ZHANG2012}
Zhang H, Schneider T, Wheeler-Kingshott CA, Alexander DC.
\newblock {NODDI}: Practical in vivo neurite orientation dispersion and density
  imaging of the human brain.
\newblock NeuroImage 2012;61:1000--1016.

\bibitem[{Lampinen et~al.(2017)Bj{\"o}rn Lampinen and Filip Szczepankiewicz and
  Johan M{\aa}rtensson and Danielle van Westen and Pia C. Sundgren and Markus
  Nilsson}]{LAMPINEN2017a}
Lampinen B, Szczepankiewicz F, M{\aa}rtensson J, van Westen D, Sundgren PC,
  Nilsson M.
\newblock Neurite density imaging versus imaging of microscopic anisotropy in
  diffusion MRI: A model comparison using spherical tensor encoding.
\newblock NeuroImage 2017;147:517--531.

\bibitem[{Hutchinson et~al.(2017)Elizabeth B. Hutchinson and Alexandru V. Avram
  and M. Okan Irfanoglu and C. Guan Koay and Alan S. Barnett and Michal E.
  Komlosh and Evren {\"O}zarslan and Susan C. Schwerin and Sharon L. Juliano
  and Carlo Pierpaoli}]{HUTCHINSON2017}
Hutchinson EB, Avram AV, Irfanoglu MO, Koay CG, Barnett AS, Komlosh ME, et~al.
\newblock Analysis of the effects of noise, DWI sampling, and value of assumed
  parameters in diffusion {MRI} models.
\newblock Magnetic Resonance in Medicine 2017;78(5):1767--1780.

\bibitem[{Jelescu et~al.(2015)Ileana O. Jelescu and Jelle Veraart and Vitria
  Adisetiyo and Sarah S. Milla and Dmitry S. Novikov and Els
  Fieremans}]{JELESCU2015a}
Jelescu IO, Veraart J, Adisetiyo V, Milla SS, Novikov DS, Fieremans E.
\newblock One diffusion acquisition and different white matter models: How does
  microstructure change in human early development based on {WMTI} and {NODDI}.
\newblock NeuroImage 2015;107:242--256.

\bibitem[{Novikov et~al.(2018)Dmitry S. Novikov and Jelle Veraart and Ileana O.
  Jelescu and Els Fieremans}]{NOVIKOV2018}
Novikov DS, Veraart J, Jelescu IO, Fieremans E.
\newblock Rotationally-invariant mapping of scalar and orientational metrics of
  neuronal microstructure with diffusion {MRI}.
\newblock NeuroImage 2018;174:518 -- 538.

\bibitem[{Reisert et~al.(2017)Marco Reisert and Elias Kellner and Bibek Dhital
  and J{\"u}rgen Hennig and Valerij G. Kiselev}]{REISERT2017}
Reisert M, Kellner E, Dhital B, Hennig J, Kiselev VG.
\newblock Disentangling micro from mesostructure by diffusion {MRI}: A
  {Bayesian} approach.
\newblock NeuroImage 2017;147(Supplement C):964 -- 975.

\bibitem[{Stejskal and Tanner(1965)E. O. Stejskal and T. E.
  Tanner}]{STEJSKAL&TANNER1965}
Stejskal EO, Tanner TE.
\newblock Spin diffusion measurements: spin echoes in the presence of a
  time‐dependent field gradient.
\newblock The Journal of Chemical Physics 1965;42:288--292.

\bibitem[{Jones(2004)Derek K. Jones}]{JONES2004}
Jones DK.
\newblock The effect of gradient sampling schemes on measures derived from
  diffusion tensor {MRI}: A {Monte} {Carlo} study.
\newblock Magnetic Resonance in Medicine 2004;51:807--815.

\bibitem[{Alexander(2008)Daniel C. Alexander}]{ALEXANDER2008}
Alexander DC.
\newblock A General Framework for Experiment Design in Diffusion {MRI} and Its
  Application in Measuring Direct Tissue-Microstructure Features.
\newblock Magnetic Resonance in Medicine 2008;60:439--448.

\bibitem[{Shemesh et~al.(2015)Noam Shemesh and Sune N. Jespersen and Daniel C.
  Alexander and Yoram Cohen and Ivana Drobnjac and Tim B. Dyrby and Jurgen
  Finterbusch and Martin A. Koch and Tristan Kuder and Fredrik Laun and Marco
  Lawrenz and Henrik Lundell and Partha P. Mitra and Markus Nilsson and Evren
  \"Ozarslan and Daniel Topgaard and Carl-Fredrik Westin}]{SHEMESH2015}
Shemesh N, Jespersen SN, Alexander DC, Cohen Y, Drobnjac I, Dyrby TB, et~al.
\newblock Conventions and Nomenclature for {Double} {Diffusion} {Encoding}
  {NMR} and {MRI}.
\newblock Magnetic Resonance in Medicine 2015;75:82--87.

\bibitem[{Cory et~al.(1990)D. G. Cory and A. N. Garroway and J. B.
  Miller}]{CORY1990}
Cory DG, Garroway AN, Miller JB.
\newblock Applications of spin transport as a probe of local geometry.
\newblock Polymer Preprints 1990;31:149.

\bibitem[{Shemesh et~al.(2010)Noam Shemesh and Evren \"Ozarslan and Michal E.
  Komlosh and Peter J. Basser and Yoram Cohen}]{SHEMESH2010a}
Shemesh N, \"Ozarslan E, Komlosh ME, Basser PJ, Cohen Y.
\newblock From single-pulsed field gradient to double-pulsed field gradient
  {MR}: gleaning new microstructural information and developing new forms of
  contrast in {MRI}.
\newblock NMR in Biomedicine 2010;23:757--780.

\bibitem[{\"Ozarslan et~al.(2009)Evren \"Ozarslan and Noam Shemesh and Peter J.
  Basser}]{OZARSLAN2009a}
\"Ozarslan E, Shemesh N, Basser PJ.
\newblock A general framework to quantify the effect of restricted diffusion on
  the {NMR} signal with applications to double pulsed field gradient {NMR}
  experiments.
\newblock The Journal of Chemical Physics 2009;130:104702/1--104702/9.

\bibitem[{Jespersen et~al.(2013)Sune N{\o}rh{\o}j Jespersen and Henrik Lundell
  and Casper Kaae S{\o}nderby and Tim B. Dyrby}]{JESPERSEN2013}
Jespersen SN, Lundell H, S{\o}nderby CK, Dyrby TB.
\newblock Orientationally invariant metrics of apparent compartment
  eccentricity from double pulsed field gradient diffusion experiments.
\newblock NMR in Biomedicine 2013;26:1647--1662.

\bibitem[{Benjamini et~al.(2014)Dan Benjamini and Michal E. Komlosh and Peter
  J. Basser and Uri Nevo}]{BENJAMINI2014}
Benjamini D, Komlosh ME, Basser PJ, Nevo U.
\newblock Nonparametric pore size distribution using d-{PFG}: Comparison to
  s-{PFG} and migration to {MRI}.
\newblock Journal of Magnetic Resonance 2014;246:36--45.

\bibitem[{Ianu\c{s} et~al.(2016)Andrada Ianu\c{s} and Ivana Drobnjak and Daniel
  C. Alexander}]{IANUS2016}
Ianu\c{s} A, Drobnjak I, Alexander DC.
\newblock Model-based estimation of microscopic anisotropy using diffusion
  {MRI}: a simulation study.
\newblock NMR in Biomedicine 2016;29:672--685.

\bibitem[{Jespersen(2011)Sune N{\o}rh{\o}j Jespersen}]{JESPERSEN2011}
Jespersen SN.
\newblock Equivalence of double and single wave vector diffusion contrast at
  low diffusion weighting.
\newblock NMR in Biomedicine 2011;25:813--818.

\bibitem[{Coelho et~al.(2017)Santiago Coelho and Leandro Beltrachini and Jose
  M. Pozo and Alejandro F. Frangi}]{COELHO2017}
Coelho S, Beltrachini L, Pozo JM, Frangi AF.
\newblock {Double} {Diffusion} {Encoding} vs {Single} {Diffusion} {Encoding} in
  Parameter Estimation of Biophysical Models in {Diffusion-Weighted} {MRI}.
\newblock In: Proceedings of the International Society of Magnetic Resonance in
  Medicine Wiley; 2017. .

\bibitem[{Fieremans et~al.(2018)Els Fieremans and Jelle Veraart and Benjamin
  Ades-Aron and Filip Szczepankiewicz and Markus Nilsson and Dmitry S
  Novikov}]{FIEREMANS2018}
Fieremans E, Veraart J, Ades-Aron B, Szczepankiewicz F, Nilsson M, Novikov DS.
\newblock Effects of combining linear with spherical tensor encoding on
  estimating brain microstructural parameters.
\newblock In: Proceedings of the International Society of Magnetic Resonance in
  Medicine Wiley; 2018. .

\bibitem[{Dhital et~al.(2018)Bibek Dhital and Marco Reisert and Elias Kellner
  and Valerij G. Kiselev}]{DHITAL2018}
Dhital B, Reisert M, Kellner E, Kiselev VG.
\newblock Diffusion Weighting with linear and planar encoding solves degeneracy
  in parameter estimation.
\newblock In: Proceedings of the International Society of Magnetic Resonance in
  Medicine Wiley; 2018. .

\bibitem[{Jespersen et~al.(2010)Sune N{\o}rh{\o}j Jespersen and Carsten R.
  Bjarkam and Jens R. Nyengaard and M. Mallar Chakravarty and Brian Hansen and
  Thomas Vosegaard and Leif {\O}stergaard and Dmitriy Yablonskiy and Niels Chr.
  Nielsen and Peter Vestergaard-Poulsen}]{JESPERSEN2010}
Jespersen SN, Bjarkam CR, Nyengaard JR, Chakravarty MM, Hansen B, Vosegaard T,
  et~al.
\newblock Neurite density from magnetic resonance diffusion measurements at
  ultrahigh field: {Comparison} with light microscopy and electron microscopy.
\newblock NeuroImage 2010;49:205--216.

\bibitem[{Alexander et~al.(2010)Daniel C. Alexander and Penny L. Hubbard and
  Matt G. Hall and Elizabeth A. Moore and Maurice Ptito and Geoff J.M. Parker
  and Tim B. Dyrby}]{ALEXANDER2010}
Alexander DC, Hubbard PL, Hall MG, Moore EA, Ptito M, Parker GJM, et~al.
\newblock Orientationally invariant indices of axon diameter and density from
  diffusion {MRI}.
\newblock NeuroImage 2010;52:1374--1389.

\bibitem[{Dhital et~al.(2017)Bibek Dhital and Elias Kellner and Valerij G.
  Kiselev and Marco Reisert}]{DHITAL2017}
Dhital B, Kellner E, Kiselev VG, Reisert M.
\newblock The absence of restricted water pool in brain white matter.
\newblock NeuroImage 2017;In press.

\bibitem[{Tax et~al.(2018)Chantal Tax and Filip Szczepankiewicz and Markus
  Nilsson and Derek Jones}]{TAX2018}
Tax C, Szczepankiewicz F, Nilsson M, Jones D.
\newblock The Dot… wherefore art thou? Search for the isotropic restricted
  diffusion compartment in the brain with spherical tensor encoding and strong
  gradients.
\newblock In: Proceedings of the International Society of Magnetic Resonance in
  Medicine Wiley; 2018. .

\bibitem[{Fieremans et~al.(2012)Els Fieremans and Jens H. Jensen and Joseph A.
  Helpern ans Sungheon Kim and Robert I. Grossman and Matilde Inglese and
  Dmitry S. Novikov}]{FIEREMANS2012}
Fieremans E, Jensen JH, ans Sungheon~Kim JAH, Grossman RI, Inglese M, Novikov
  DS.
\newblock Diffusion distinguishes between axonal loss and demyelination in
  brain white matter.
\newblock In: Proceedings of the International Society of Magnetic Resonance in
  Medicine Wiley; 2012. .

\bibitem[{Jelescu et~al.(2016)Ileana O. Jelescu and Magdalena Zurek and
  Kerryanne V. Winters and Jelle Veraart and Anjali Rajaratnam and Nathanael S.
  Kim and James S. Babb and Timothy M. Shepherd and Dmitry S. Novikov and
  Sungheon G. Kim and Els Fieremans}]{JELESCU2016}
Jelescu IO, Zurek M, Winters KV, Veraart J, Rajaratnam A, Kim NS, et~al.
\newblock \textit{In vivo} quantification of demyelination and recovery using
  compartment-specific diffusion {MRI} metrics validated by electron
  microscopy.
\newblock NeuroImage 2016;132:104--114.

\bibitem[{Budde and Frank(2010)Matthew D. Budde and Joseph A.
  Frank}]{BUDDE2010}
Budde MD, Frank JA.
\newblock Neurite beading is sufficient to decrease the apparent diffusion
  coefficient after ischemic stroke.
\newblock Proceedings of the National Academy of Sciences
  2010;107(32):14472--14477.

\bibitem[{Jelescu et~al.(2016)Ileana O. Jelescu and Jelle Veraart and Els
  Fieremans and Dmitry S. Novikov}]{JELESCU2015b}
Jelescu IO, Veraart J, Fieremans E, Novikov DS.
\newblock Degeneracy in model parameter estimation for multi-compartmental
  diffusion in neuronal tissue.
\newblock NMR in Biomedicine 2016;29:33--47.

\bibitem[{Hansen et~al.(2016)Brian Hansen and Noam Shemesh and Sune
  N{\o}rh{\o}j Jespersen}]{HANSEN2016}
Hansen B, Shemesh N, Jespersen SN.
\newblock Fast imaging of mean, axial and radial diffusion kurtosis.
\newblock NeuroImage 2016;142:381--393.

\bibitem[{Jespersen et~al.(2017)Sune N{\o}rh{\o}j Jespersen and Jonas Lynge
  Olesen and Brian Hansen and Noam Shemesh}]{JESPERSEN2017}
Jespersen SN, Olesen JL, Hansen B, Shemesh N.
\newblock Diffusion time dependence of microstructural parameters in fixed
  spinal cord.
\newblock NeuroImage 2017;In press.

\bibitem[{Abramowitz et~al.(1972)M. Abramowitz and I. A. Stegun and (Firm)
  Knovel}]{ABRAMOWITZ1972}
Abramowitz M, Stegun IA, Knovel F.
\newblock Handbook of Mathematical Functions with Formulas, Graphs, and
  Mathematical Tables.
\newblock National Bureau of Standards Applied mathematics series 55; 1972.

\bibitem[{Hansen et~al.(2017)Brian Hansen and Ahmad R. Khan and Noam Shemesh
  and Torben E. Lund and Ryan Sangill and Simon F. Eskildsen and Leif
  {\O}stergaard and Sune N{\o}rh{\o}j Jespersen}]{HANSEN2017}
Hansen B, Khan AR, Shemesh N, Lund TE, Sangill R, Eskildsen SF, et~al.
\newblock White matter biomarkers from fast protocols using axially symmetric
  diffusion kurtosis imaging.
\newblock NMR in Biomedicine 2017;30:1--17.

\bibitem[{Westin et~al.(2014){Carl-Fredrik} Westin and Filip Szczepankiewicz
  and Ofer Pasternak and Evren \"Ozarslan and Daniel Topgaard and Hans Knutsson
  and Markus Nilsson}]{WESTIN2014}
Westin C, Szczepankiewicz F, Pasternak O, \"Ozarslan E, Topgaard D, Knutsson H,
  et~al.
\newblock Measurement Tensors in Diffusion {MRI}: Generalizing the Concept of
  Diffusion Encoding.
\newblock In: Medical Image Computing and Computer-Assisted Intervention
  (MICCAI), vol. 8675 Springer; 2014. p. 209--216.

\bibitem[{Westin et~al.(2016)Carl-Fredrik Westin and Hans Knutsson and Ofer
  Pasternak and Filip Szczepankiewicz and Evren \"Ozarslan and Danielle van
  Westen and Cecilia Mattisson and Mats Bogren and Lauren J. O'Donnell and
  Marek Kubicki and Daniel Topgaard and Markus Nilsson}]{WESTIN2016}
Westin CF, Knutsson H, Pasternak O, Szczepankiewicz F, \"Ozarslan E, van Westen
  D, et~al.
\newblock q-space trajectory imaging for multidimensional diffusion {MRI} of
  the human brain.
\newblock NeuroImage 2016;135:345--362.

\bibitem[{Itin and Hehl(2013)Yakov Itin and Friedrich W. Hehl}]{ITIN2013}
Itin Y, Hehl FW.
\newblock The constitutive tensor of linear elasticity: Its decompositions,
  Cauchy relations, null Lagrangians, and wave propagation.
\newblock Journal of Mathematical Physics 2013;54(6042903).

\bibitem[{Neuman(1974)C. H. Neuman}]{NEUMAN1974}
Neuman CH.
\newblock Spin echo of spins diffusing in a bounded medium.
\newblock The Journal of Chemical Physics 1974;60:4508--4511.

\bibitem[{Lebedev and Laikov(1999)V.I. Lebedev and D.N. Laikov}]{LEBEDEV1999}
Lebedev VI, Laikov DN.
\newblock A quadrature formula for the sphere of the 131st algebraic order of
  accuracy.
\newblock Doklady Mathematics 1999;59:477--481.

\bibitem[{Zhang et~al.(2011)Hui Zhang and Penny L. Hubbard and Geoff J.M.
  Parker and Daniel C. Alexander}]{ZHANG2011}
Zhang H, Hubbard PL, Parker GJM, Alexander DC.
\newblock Axon diameter mapping in the presence of orientation dispersion with
  diffusion {MRI}.
\newblock NeuroImage 2011;56:1301--1315.

\bibitem[{Gudbjartsson and Patz(1995)H\'{a}kon Gudbjartsson and Samuel
  Patz}]{GUDBJARTSSON1995}
Gudbjartsson H, Patz S.
\newblock The {Rician} distribution of noisy {MRI} data.
\newblock Magnetic Resonance in Medicine 1995;34(6):910--914.

\bibitem[{Koay et~al.(2012)Cheng Guan Koay and Evren \"Ozarslan and Kevin M.
  Johnson and M. Elizabeth Meyerand}]{KOAY2012}
Koay CG, \"Ozarslan E, Johnson KM, Meyerand ME.
\newblock Sparse and optimal acquisition design for diffusion {MRI} and beyond.
\newblock Medical Physics 2012;39:2499--2511.

\bibitem[{Reisert et~al.(2018)Marco Reisert and Valerij G. Kiselev and Bibek
  Dhital}]{REISERT2018}
Reisert M, Kiselev VG, Dhital B.
\newblock A Unique Analytical Solution of the White Matter Standard Model using
  Linear and Planar Encodings.
\newblock Preprint arXiv 2018;1808.04389v1.

\bibitem[{Veraart et~al.(2017)Jelle Veraart and Dmitry S. Novikov and Els
  Fieremans}]{VERAART2017}
Veraart J, Novikov DS, Fieremans E.
\newblock {TE} dependent {Diffusion} {Imaging} ({TEdDI}) distinguishes between
  compartmental {T2} relaxation times.
\newblock NeuroImage 2017;In press.

\bibitem[{Tariq et~al.(2016)Maira Tariq and Torben Schneider and Daniel C.
  Alexander and Claudia A. Gandini {Wheeler-Kingshott} and Hui
  Zhang}]{TARIQ2016}
Tariq M, Schneider T, Alexander DC, {Wheeler-Kingshott} CAG, Zhang H.
\newblock {Bingham–NODDI}: Mapping anisotropic orientation dispersion of
  neurites using diffusion {MRI}.
\newblock NeuroImage 2016;133:207 -- 223.

\end{thebibliography}




\end{document}